\documentclass{SciPost}

\usepackage{graphicx}
\usepackage{dcolumn}
\usepackage{bm}
\usepackage{braket}
\usepackage{amsmath}
\usepackage{amssymb}
\usepackage{amsfonts}
\usepackage{latexsym}
\usepackage{mathtools}
\usepackage{bbm}
\usepackage{color}
\usepackage[normalem]{ulem}
\usepackage {comment}

\hypersetup{
    colorlinks,
    linkcolor={red!50!black},
    citecolor={blue!50!black},
    urlcolor={blue!80!black}
}

\usepackage[bitstream-charter]{mathdesign}
\DeclareSymbolFont{usualmathcal}{OMS}{cmsy}{m}{n}
\DeclareSymbolFontAlphabet{\mathcal}{usualmathcal}

\fancypagestyle{SPstyle}{
\fancyhf{}
\lhead{\colorbox{scipostblue}{\bf \color{white} ~SciPost Physics }}
\rhead{{\bf \color{scipostdeepblue} ~Submission }}

\fancyfoot[C]{\textbf{\thepage}}
}

\begin{document}

\pagestyle{SPstyle}

\begin{center}{\Large \textbf{\color{scipostdeepblue}{
A sine-square deformation approach to quantum critical points in one-dimensional systems\\
}}}\end{center}

\begin{center}\textbf{
Y. Miyazaki\textsuperscript{1$\star$},
S. Tanigawa\textsuperscript{2},
G. Marmorini\textsuperscript{2},
N. Furukawa\textsuperscript{3}, and 
D. Yamamoto\textsuperscript{2,4}
}\end{center}

\begin{center}
{\bf 1} Computational Materials Science Research Team, RIKEN Center for Computational Science (R-CCS), Hyogo 650-0047, Japan
\\
{\bf 2} Department of Physics, College of Humanities and Sciences, Nihon University, Tokyo 156-8550, Japan
\\
{\bf 3} Department of Physical Science, Aoyama Gakuin University, Kanagawa 252-5258, Japan
\\
{\bf 4} 
Quantum Many-Body Dynamics Research Team, RIKEN Center for Quantum Computing (RQC), Wako, Saitama 351-0198, Japan

$\star$ \href{mailto:email1}{\small yuki.miyazaki.kj@riken.jp}\,
\end{center}

\section*{\color{scipostdeepblue}{Abstract}}
\textbf{\boldmath{
We propose a method to determine the quantum phase boundaries of one-dimensional systems using sine-square deformation (SSD).
Based on the proposition, supported by several exactly solved cases though not proven in full generality, that ``if a one-dimensional system is gapless, then the expectation value of any local observable in the ground state of the Hamiltonian with SSD exhibits translational symmetry in the thermodynamic limit," we determine the quantum critical point as the location where a local observable becomes site-independent, identified through finite-size scaling analysis.
As case studies, we consider two models: the antiferromagnetic Ising chain in mixed transverse and longitudinal magnetic fields with nearest-neighbor and long-range interactions. 
We calculate the ground state of these Hamiltonians with SSD using the density-matrix renormalization-group algorithm and evaluate the local transverse magnetization.
For the nearest-neighbor model, we show that the quantum critical point can be accurately estimated by our procedure with systems of up to 84 sites, or even smaller, in good agreement with results from the literature. For the long-range model, we find that the phase boundary between the antiferromagnetic and paramagnetic phases is slightly shifted relative to the nearest-neighbor case, leading to a reduced region of antiferromagnetic order.
Moreover, we propose an experimental procedure to implement the antiferromagnetic $J_1$-$J_2$ Ising couplings with SSD using Rydberg atom arrays in optical tweezers, which can be achieved within a very good approximation. 
We show that our protocol can control the ratio $J_2/J_1$ over a wide range, and realize a well-approximated SSD of the nearest-neighbor model.
Because multiple independent scaling conditions naturally emerge, our approach enables precise determination of quantum critical points and possibly even the extraction of additional critical phenomena, such as critical exponents, from relatively small system sizes. 
}}

\vspace{\baselineskip}

\noindent\textcolor{white!90!black}{%
\fbox{\parbox{0.975\linewidth}{%
\textcolor{white!40!black}{\begin{tabular}{lr}%
  \begin{minipage}{0.6\textwidth}%
    {\small Copyright attribution to authors. \newline
    This work is a submission to SciPost Physics. \newline
    License information to appear upon publication. \newline
    Publication information to appear upon publication.}
  \end{minipage} & \begin{minipage}{0.4\textwidth}
    {\small Received Date \newline Accepted Date \newline Published Date}%
  \end{minipage}
\end{tabular}}
}}
}




\vspace{10pt}
\noindent\rule{\textwidth}{1pt}
\tableofcontents
\noindent\rule{\textwidth}{1pt}
\vspace{10pt}


\section{Introduction}
\label{sec:intro}

Quantum phase transitions are among the most intriguing phenomena in quantum many-body physics, offering deep insights into universality, emergent properties, and low-energy excitations.
Since a true quantum phase transition occurs only in the thermodynamic limit, numerical analyses are typically performed for finite-size systems and rely on finite-size scaling to estimate phase boundaries and to characterize the nature of different phases. In finite-size calculations, the choice of  boundary conditions plays a crucial role as translational symmetry strongly affects the physical properties of the system.
Periodic boundary conditions (PBCs) preserve translational symmetry and make the system spatially uniform, whereas open boundary conditions (OBCs) break translational symmetry at the edges, producing boundary effects that differ from bulk behavior.
From the perspective of studying bulk properties and  phase transitions, it is often preferable to adopt PBCs, since they minimize boundary effects and make it easier to capture intrinsic bulk behavior.

Nevertheless, OBCs remain important for several reasons. First, they allow one to access boundary-related phenomena such as edge states and topological excitations \cite{SSH,topo,edge,Haldane1,Haldane2}, which cannot appear under PBCs. Second, OBCs are computationally advantageous for many finite tensor-network algorithms based on variational wavefunction representations such as matrix product states (MPS) and projected entangled pair states (PEPS) where open boundaries simplify the implementation and improve convergence \cite{MPS1,MPS2,PEPS1,PEPS2,PEPS3}. Finally, in the context of quantum simulation, physical systems such as cold atomic gases in optical lattice \cite{coldatom1,coldatom2,coldatom3}, trapped-ion arrays \cite{trappedion1,trappedion2,trappedion3}, and Rydberg atom arrays \cite{Rydberg1,Rydberg2,Rydberg3,Rydberg4,Rydberg5,Rydberg6} naturally realize open boundaries due to their finite spatial extent and confinement. In the latter two cases, it is essential to suppress unwanted boundary effects as effectively as possible while retaining the open geometry.

\begin{figure}[t]
    \centering
    \includegraphics[width=12cm]{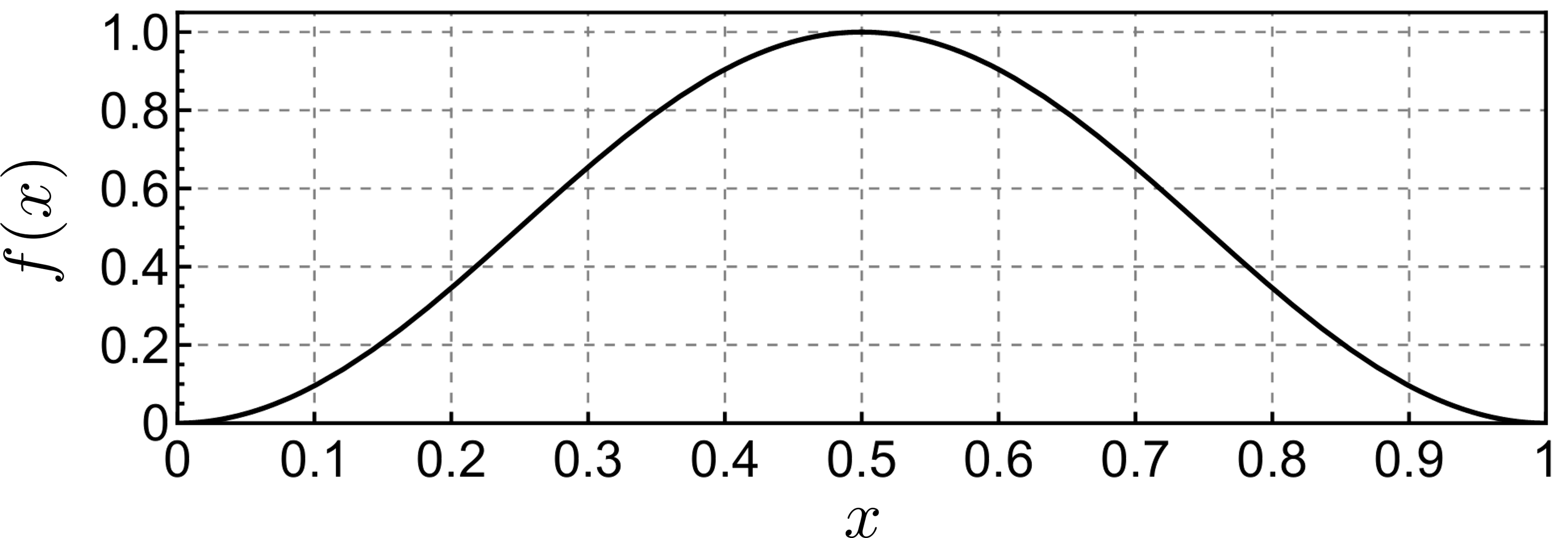}
    \caption{Continuous form of the modulation function $f_L(i)$ in the sine-square deformation, plotted as $f(x)=\sin^2(\pi x)$ with $x=(i-1/2)/L$. The function smoothly decreases from unity at the center to zero at both edges.}
   \label{shape_SSD}
\end{figure}

Vekić and White proposed the concept of smooth boundary conditions for one-dimensional systems, in which the parameters of an open-boundary Hamiltonian are smoothly tuned to zero near the edges \cite{SBC1}.
They demonstrated that such smooth boundaries effectively reduce finite-size effects and enable reliable extrapolation to the thermodynamic limit using relatively small systems.
Following their work, various forms of smooth boundary conditions have been explored and extended to higher-dimensional models \cite{SBC2,SBC3,SBC4}. 
Among these approaches, the sine-square deformation (SSD) \cite{SSD1,SSD2,SSD3,SSD4,SSD5,SSD6,SSD7} was introduced as a particularly effective type of smooth boundary modulation. In SSD, the local energy scale of the Hamiltonian is modulated by a site-dependent function 
\begin{equation}    f_L(i)=\sin^2\left[\frac{\pi}{L}\left(i-\frac{1}{2}\right)\right],
    \label{SSD}
\end{equation}
where $L$ is the system size and $i=1,2,\cdots,L$ is the site index. 
The continuous form of this function, $f(x) = \sin^2(\pi x)$ with $x = (i - 1/2)/L$, is shown in Fig. \ref{shape_SSD}.
The function reaches its maximum value $1$ at the center of the chain $i=(L+1)/2$, and continuously decreases to zero at the edges, thereby smoothly suppressing boundary effects while retaining open boundaries.

Previous studies have shown that SSD can almost completely eliminate the spatial oscillations of local observables that typically appear in open chains. This behavior has been confirmed in a variety of models, including the tight-binding chain \cite{SSD1}, the $XXZ$ chain \cite{SSD2}, the extended Hubbard model \cite{SBC4}, and the Kondo lattice model \cite{SSD4}.
Furthermore, the central charge obtained from the scaling of the entanglement entropy, following the Calabrese-Cardy's formula \cite{c_EE}, has been found to be identical for systems with SSD and those with PBCs  in the $XXZ$ chain, the antiferromagnetic $J_1$-$J_2$ chain, and two-leg ladder model \cite{SSD2}.
The SSD framework has also been extended to two-dimensional systems, such as the tight-binding models on the square lattice \cite{SSD_2D_1}, the Heisenberg model on the square \cite{SSD_2D_2}, triangular \cite{SSD_2D_2}, and kagome lattice \cite{SSD_2D_3,SSD_2D_4}, suggesting that the same mechanism can work beyond one dimension. 

SSD has a firm theoretical foundation in conformal field theory (CFT) \cite{SSD_CFT1,SSD_CFT2,SSD_CFT3,SSD_CFT4,SSD_CFT5,SSD_CFT6,SSD_CFT7}. In a broad class of one-dimensional critical systems described by CFT, it has been proven that the ground state of the Hamiltonian with SSD is an exact eigenstate of the Hamiltonian with PBCs. 
In particular, for free-fermion models such as the tight-binding chain and the transverse-field $XY$ (Ising) chain \cite{SSD_CFT1}, this equivalence holds exactly even for finite system sizes, not merely in the thermodynamic or conformal limit. These results indicate that the correspondence between SSD and PBC ground states can be stronger than what is implied by CFT alone.

In this work, we present a method for determining quantum critical points (QCPs) in finite quantum systems by exploiting the unique properties of the SSD. The central idea is motivated by the observation, supported by several exactly solved models, that for many gapless systems the expectation value of local observables in the ground state of the Hamiltonian with SSD tends to become translationally invariant in the thermodynamic limit.
By taking advantage of this property, we identify the QCP as the parameter value at which a local observable becomes spatially uniform, and we determine it quantitatively through finite-size scaling. Because any local observable tends to become uniform at criticality, multiple independent scaling conditions naturally emerge, which significantly enhances the accuracy of the estimation even for relatively small systems. This feature naturally matches the scale and controllability of current artificial quantum platforms, such as noisy intermediate-scale quantum (NISQ) devices \cite{NISQ1,NISQ2,NISQ3} and quantum simulators \cite{coldatom1,coldatom2,coldatom3,trappedion1,trappedion2,trappedion3,Rydberg1,Rydberg2,Rydberg3,Rydberg4,Rydberg5,Rydberg6} based on ultracold atoms, trapped ions, or Rydberg atom arrays, where the accessible system size is still limited but high-precision measurements of local observables are possible. Moreover, the present approach may be extended to study critical phenomena beyond locating QCPs, including the estimation of critical exponents and scaling behavior. This method is particularly suited for transitions involving at least one gapped phase, while its applicability to gapless--gapless transitions remains an open question for future investigation. 

As a demonstration of the proposed approach, we apply it to the antiferromagnetic spin-$1/2$ Ising chains in mixed transverse and longitudinal magnetic fields, which serves as a minimal model for exploring field-induced quantum phase transitions. To examine the generality of the method, we consider both the nearest-neighbor and long-range interaction cases. The ground states of these Hamiltonians with SSD are calculated using the density-matrix renormalization-group (DMRG) algorithm \cite{DMRG1,DMRG2,DMRG3}, and the site dependence of the local transverse magnetization is analyzed to determine the quantum phase boundaries.
In addition to numerical simulations, we also discuss a possible implementation of SSD in a Rydberg-atom quantum simulator \cite{Rydberg1,Rydberg2,Rydberg3,Rydberg4,Rydberg5,Rydberg6}, illustrating how an effective SSD-like spatial modulation of interaction strength can be realized in a realistic setup. Although this part is supplementary to the main theoretical analysis, it demonstrates the experimental feasibility of SSD-based Hamiltonians in current quantum simulation platforms.

The remainder of this paper is organized as follows. 
Section \ref{sec2} introduces the theoretical framework of the proposed method for estimating QCPs, which is based on the relationship between gapless states in one-dimensional systems and SSD.
We apply this procedure to two models: (i) the mixed-field antiferromagnetic spin-$1/2$ Ising chain with nearest-neighbor couplings, and (ii) the corresponding model with long-range interactions.
We also discuss the applicability of this approach to the analysis of the universality class of the transitions.
In Sec. \ref{sec3}, we show that the SSD of the $J_1$-$J_2$ Ising chain (including the case of very small $J_2/J_1$) can be implemented in a quantum simulator consisting of Rydberg atom arrays in optical tweezers.
Section \ref{sec4} concludes the paper.

\section{SSD approach to quantum criticality}
\label{sec2}

\subsection{Method and models}

When a system is described by a CFT, the ground state of the Hamiltonian with PBC becomes an exact eigenstate of the SSD Hamiltonian~\cite{SSD_CFT1}. 
In several important cases where the CFT is associated with the Kac-Moody algebra, this state is, futhermore, the ground state of the SSD Hamiltonian (note that it is not necessarily nondegenerate)~\cite{SSD_CFT1}.

Motivated by this correspondence in the continuum limit, we now turn to discrete lattice models and investigate how similar behavior emerges in their thermodynamic limit. In particular, we develop a method to characterize the criticality of such systems by utilizing the spatial structure of local observables in the SSD ground state. In general, the translational symmetry is broken under OBCs, and therefore the expectation values of local observables at sites $i$ and $j$ do not coincide even when the SSD is applied, except for symmetry-related points. However, when the parameters of the Hamiltonian are tuned such that the ground state becomes critical, particularly when it corresponds to the case where the SSD ground state coincides with that under PBCs, as in the CFT description, the difference between local observables at different sites is expected to diminish as the system size $L$ increases:
\begin{eqnarray}
\braket{\psi_{\rm SSD}|\hat{O}_i|\psi_{\rm SSD}}-\braket{\psi_{\rm SSD}|\hat{O}_{j}|\psi_{\rm SSD}} \rightarrow 0 \quad (i \neq j,\ {L\rightarrow\infty}),
    \label{prop}
\end{eqnarray} 
where $\ket{\psi_{\mathrm{SSD}}}$ denotes the ground state of the Hamiltonian with SSD and $\hat{O}_i$ is an arbitrary local observable at site $i$. Based on this property, we introduce a practical criterion for locating the quantum critical point in finite systems. The key idea is that the spatial variance of local observables in the SSD ground state provides a quantitative measure of the uniformity that becomes minimal at criticality.

\begin{figure}[b]
    \centering
    \includegraphics[width=13cm]{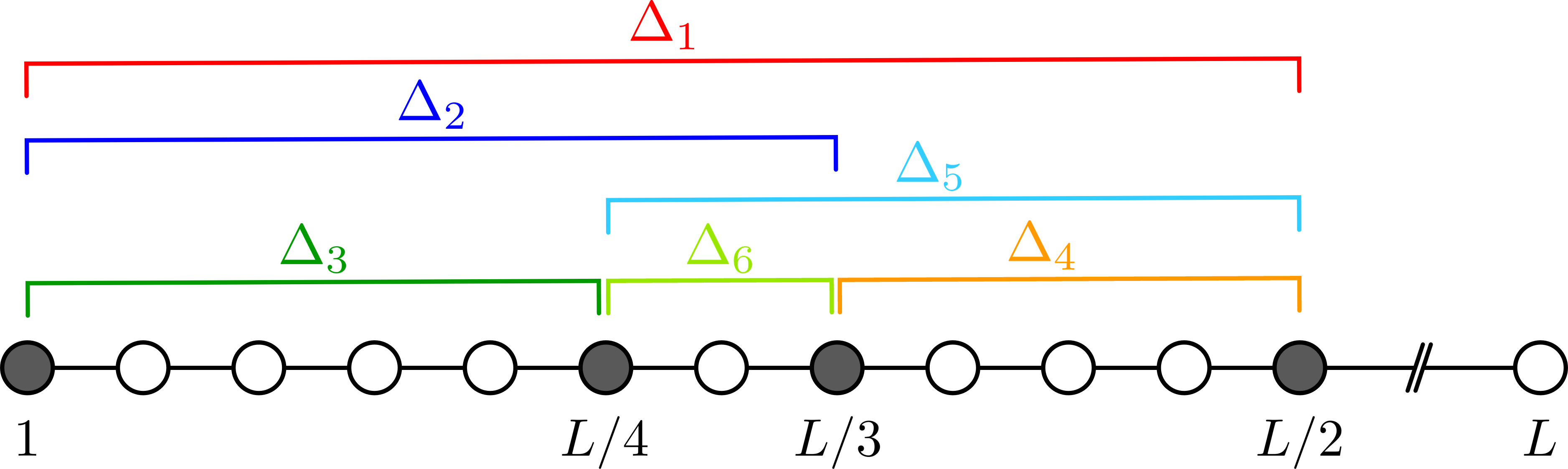}
    \caption{Schematic illustrations of the site pairs in the definitions of $\Delta_1$, $\Delta_2$, $\Delta_3$, $\Delta_4$, $\Delta_5$, and $\Delta_6$.}
   \label{def_delta}
\end{figure} 

As case studies, we test this criterion on two representative models that exhibit quantum phase transitions.
The first one is the mixed-field antiferromagnetic spin-1/2 Ising chain with nearest-neighbor couplings under OBCs, and the second is its long-range-interaction counterpart:
\begin{eqnarray}
\hat{\mathcal{H}}_{\rm NN}&=&\sum_{i=1}^{L-1}\hat{S}_i^z\hat{S}_{i+1}^z-h^x\sum_{i=1}^L \hat{S}_i^x
-h^z\sum_{i=1}^L \hat{S}_i^z
\label{H_TIM_bare}
\end{eqnarray}
and
\begin{eqnarray}
\hat{\mathcal{H}}_{\rm LR}&=&\sum_{i<j}\frac{1}{|i-j|^{6}}\hat{S}_i^z\hat{S}_{j}^z-h^x\sum_{i=1}^L \hat{S}_i^x
-h^z\sum_{i=1}^L \hat{S}_i^z,
\label{H_TIM_LR_bare}
\end{eqnarray}
where $\hat{S}_i^x$ and $\hat{S}_i^z$ are the spin-$1/2$ operators acting on site $i$, and $h^x, h^z$ denote the intensity of transverse and longitudinal fields, respectively. 
These models are known to exhibit a field-induced $\mathbb{Z}_2$ quantum phase transition between two gapped phases, namely antiferromagnetic (AFM) and paramagnetic (PM) phases~\cite{Ising_pd}.

For each model, we calculate the ground state of the Hamiltonian with SSD using the DMRG algorithm~\cite{DMRG1,DMRG2,DMRG3} and analyze the site dependence of local observables $\{\hat{O}_i\}_{i=1}^L$ to identify the QCP. The calculations are performed for system sizes $L=12$, $24$, $36$, $48$, $60$, $72$, and $84$, with the maximum bond dimension of the matrix-product state (MPS) set to 600. Specifically, we choose the transverse spin component $\hat{O}_i = \hat{S}_i^x$ and define the following quantities (see also Fig. \ref{def_delta}): 
\begin{equation}
\begin{aligned}
    \Delta_1 &\equiv \braket{\psi_{\mathrm{SSD}}|\hat{S}^x_{L/2}|\psi_{\mathrm{SSD}}} - \braket{\psi_{\mathrm{SSD}}|\hat{S}^x_{1}|\psi_{\mathrm{SSD}}}, \\
    \Delta_2 &\equiv \braket{\psi_{\mathrm{SSD}}|\hat{S}^x_{L/3}|\psi_{\mathrm{SSD}}} - \braket{\psi_{\mathrm{SSD}}|\hat{S}^x_{1}|\psi_{\mathrm{SSD}}}, \\
    \Delta_3 &\equiv \braket{\psi_{\mathrm{SSD}}|\hat{S}^x_{L/4}|\psi_{\mathrm{SSD}}} - \braket{\psi_{\mathrm{SSD}}|\hat{S}^x_{1}|\psi_{\mathrm{SSD}}}, \\
    \Delta_4 &\equiv \braket{\psi_{\mathrm{SSD}}|\hat{S}^x_{L/2}|\psi_{\mathrm{SSD}}} - \braket{\psi_{\mathrm{SSD}}|\hat{S}^x_{L/3}|\psi_{\mathrm{SSD}}}, \\
    \Delta_5 &\equiv \braket{\psi_{\mathrm{SSD}}|\hat{S}^x_{L/2}|\psi_{\mathrm{SSD}}} - \braket{\psi_{\mathrm{SSD}}|\hat{S}^x_{L/4}|\psi_{\mathrm{SSD}}}, \\
    \Delta_6 &\equiv \braket{\psi_{\mathrm{SSD}}|\hat{S}^x_{L/3}|\psi_{\mathrm{SSD}}} - \braket{\psi_{\mathrm{SSD}}|\hat{S}^x_{L/4}|\psi_{\mathrm{SSD}}}.
\end{aligned}
\label{eq:deltas}
\end{equation}
According to the proposition~(\ref{prop}), all $\Delta_n$ tend to approach zero at the QCP in the thermodynamic limit. We determine the zero points of $\{\Delta_n\}_{n=1}^6$ as functions of the system size $L$, and estimate the phase boundary in the thermodynamic limit by least-squares extrapolation using the data of finite-size systems.

Note that, to detect the phase transitions of the two models, the staggered magnetization $\sum_i (-1)^i \hat{S}_i^z$ is usually employed as an order parameter.  
In practice, one often evaluates its square average in finite systems to avoid the vanishing expectation value due to symmetry.  
In the present approach, however, any local quantity can be used, as long as its expectation value is not trivially zero, such as $\hat{O}_i = \hat{S}_i^z$ (widely employed here), two-site operators like $\hat{O}_i = \hat{S}_i^z \hat{S}_{i+1}^z$, etc.

\subsection{Nearest-neighbor model}
In this subsection, we discuss the mixed-field antiferromagnetic spin-1/2 Ising chain with nearest-neighbor interactions, $\hat{\mathcal{H}}_{\mathrm{NN}}$ [Eq.~\eqref{H_TIM_bare}].  
In the simple case of $h^z = 0$, corresponding to the transverse-field Ising model, it is well known that the system undergoes a quantum phase transition from an antiferromagnetic (AFM) to a paramagnetic (PM) phase as $h^x$ increases, with the critical point located at $h^x_{\mathrm{c}} = 0.5$. When a finite longitudinal field $h^z$ is applied, the location of the QCP shifts in the $(h^x, h^z)$ plane, following a continuous curve that eventually reaches the transition point of the classical Ising model at $(h^x_{\mathrm{c}}, h^z_{\mathrm{c}}) = (0, 1)$.

Here, we revisit this problem by applying our SSD approach, which takes advantage of the property that the SSD ground state asymptotically approaches the ground state of the system with PBCs and thereby restores translational symmetry at special parameter points. This feature enables us to determine the location of the quantum critical point and its evolution in the $(h^x,h^z)$ plane with high accuracy, even for relatively small system sizes.
The Hamiltonian with SSD is given by
\begin{equation}
\hat{\mathcal{H}}_{\rm NN}^{\rm (SSD)}=\sum_{i=1}^{L-1}f_L\left(i+\tfrac{1}{2}\right)\hat{S}_i^z\hat{S}_{i+1}^z
-h^x\sum_{i=1}^L f_L(i)\hat{S}_i^x
-h^z\sum_{i=1}^L f_L(i)\hat{S}_i^z.
\label{H_TIM}
\end{equation}

\begin{figure}[t]
    \centering
    \includegraphics[width=9cm]{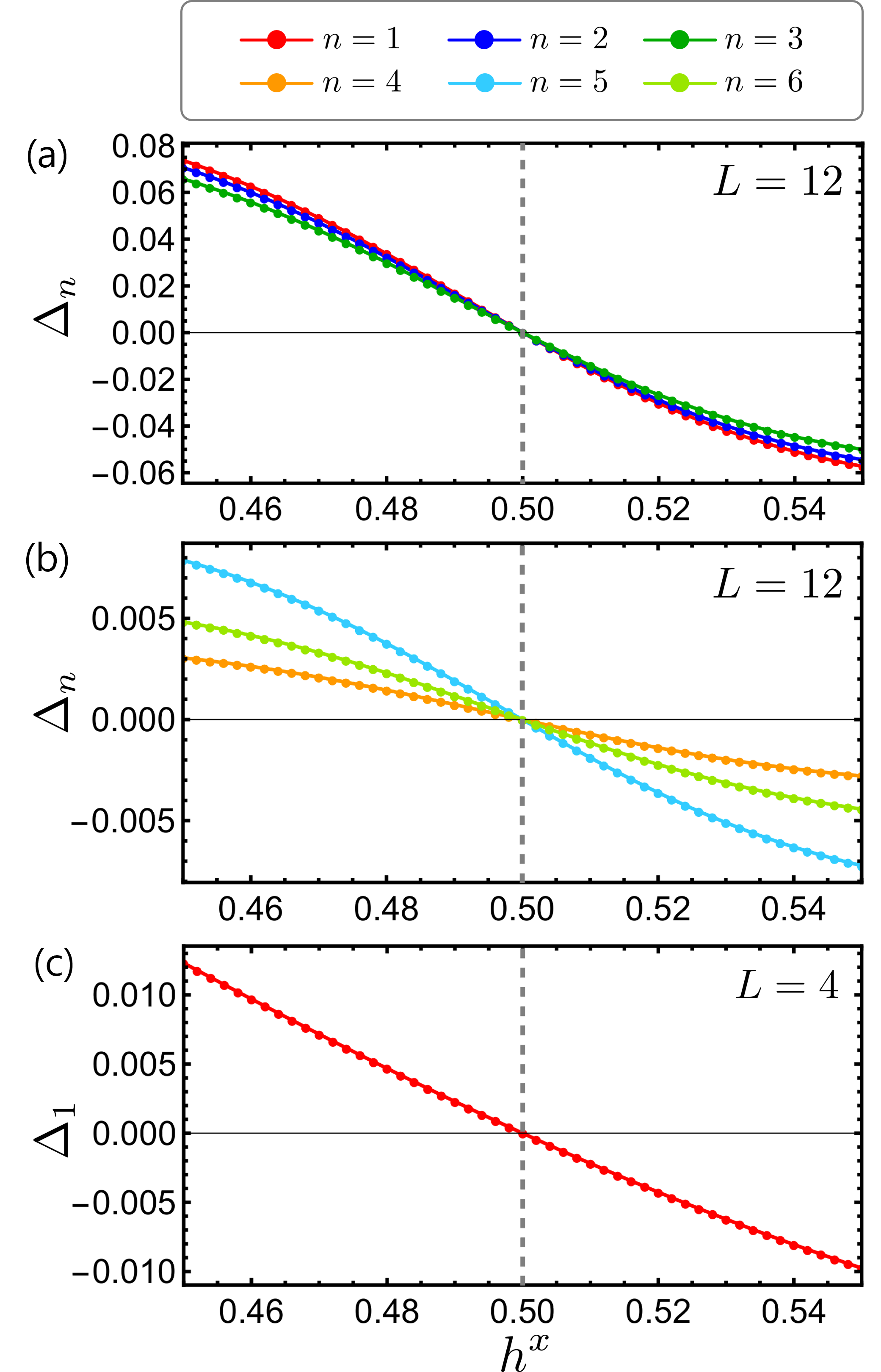}
    \caption{(a, b) $\Delta_n$ ($n=1$–$6$) as functions of $h^x$ for the ground state of the Hamiltonian~(\ref{H_TIM}) with $h^z=0$ and $L=12$. 
    (c) $\Delta_1$ for $L=4$. 
    The gray dashed line indicates the QCP.}
   \label{TIM_zero_delta}
\end{figure}

We first examine the case of $h^z=0$. Figures~\ref{TIM_zero_delta}(a) and \ref{TIM_zero_delta}(b) show the values of $\{\Delta_n\}_{n=1}^6$ as functions of $h^x$ for $L=12$. Strikingly, all $\Delta_n$ vanish simultaneously the known QCP, $h^x_{\mathrm{c}} = 0.5$, indicating that the SSD ground state becomes translationally invariant. In this model, the system is expected to become gapless at the QCP in the thermodynamic limit, where the low-energy physics is described by CFT~\cite{Ising_CFT}. However, the fact that all $\Delta_n$ vanish exactly even for $L = 12$, far from the continuum field theory limit, indicates that the correspondence between the SSD and PBC ground states holds exactly in this special case, regardless of the CFT description. Indeed, the equivalence between the SSD and PBC ground states at the QCP of the transverse Ising chain (or, more generally, of the anisotropic $XY$ spin chains in a transverse field) has been proven in Ref.~\cite{SSD_CFT1}, independently of the system size $L$. Therefore, as shown in Fig.~\ref{TIM_zero_delta}(c), the emergence of translational invariance in the SSD ground state, reflected by the vanishing of $\Delta_1$, occurs even for $L=4$ exactly at the QCP.

We next investigate the case of a finite longitudinal field $h^z > 0$. In this case, the $\Delta_n$'s are not expected to vanish completely at the QCP for finite $L$, since the exact ground-state equivalence between the SSD and PBC ground states no longer holds. Nevertheless, as the system size $L$ increases, all $\Delta_n$ tend to approach zero or change sign only near the QCP, where the system becomes gapless and is described by a CFT with central charge $c = 1/2$ \cite{Ising_mixedfield_CFT}. 
We estimate the QCP from the restoration of translational invariance in the thermodynamic limit $L\to\infty$, assuming that SSD and PBC ground states become equivalent at the QCP described by the CFT.

\begin{figure*}[t]
    \centering
    \includegraphics[width=14cm]{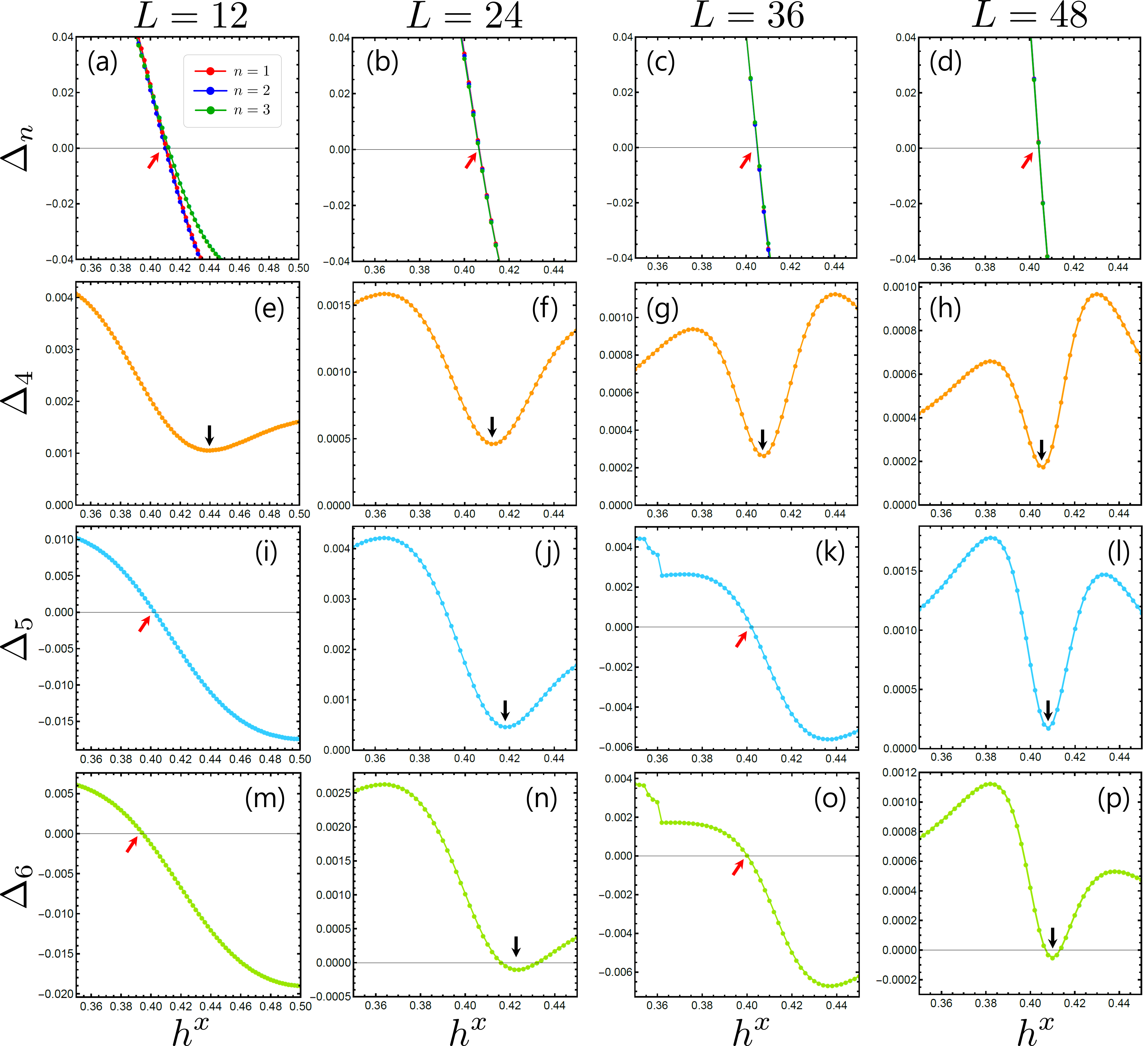}
    \caption{
$\Delta_n$ ($n=1$–$6$) for the ground state of the nearest-neighbor model with SSD [Eq.~(\ref{H_TIM})] at $h^z = 0.5$, 
    shown as functions of $h^x$.  
    Panels (a)–(p) are arranged in a $4\times4$ grid, 
    where each row corresponds to a different $\Delta_n$ 
    [$\Delta_{1,2,3}$ in the first row, $\Delta_4$ in the second, $\Delta_5$ in the third, and $\Delta_6$ in the fourth], 
    and each column corresponds to a different system size 
    ($L = 12, 24, 36,$ and $48$ from left to right).  
    Red arrows indicate the zero-crossing points, and black arrows indicate the local-minimum points adopted for the estimation of the QCP.
    }
    \label{TIM_deltas}
\end{figure*}

Figures~\ref{TIM_deltas}(a--p) show $\Delta_n$ ($n=1$--$6$) as functions of $h^x$ for $h^z = 0.5$ and several system sizes $L = 12, 24, 36,$ and $48$.  
Two distinct types of behavior are observed depending on $n$: 
$\Delta_1$, $\Delta_2$, and $\Delta_3$ change sign around the QCP, whereas $\Delta_4$ exhibits a local minimum with a value close to zero near the QCP.  
The behavior of $\Delta_5$ and $\Delta_6$ depends on the system size; they show either sign changes or local minima depending on $L$.  
These differences are attributed to the parity combinations of the reference sites, such as $L/2$, $L/3$, and $L/4$, that enter the definitions of $\Delta_n$. 
This parity dependence is naturally related to the staggered structure of antiferromagnetic correlations in the ground state, where local observables inherit an oscillatory component with period two. In finite systems, such an alternating structure can manifest itself differently for site pairs separated by odd or even distances.
For each $\Delta_n$, we denote by $(h^x_{\mathrm{cross}})_n$ the value of $h^x$ at which $\Delta_n$ crosses zero, and by $(h^x_{\mathrm{min}})_n$ the value at which it takes a local minimum. 
Despite the variations in behavior among different $\Delta_n$, both $(h^x_{\mathrm{cross}})_n$ and $(h^x_{\mathrm{min}})_n$ tend to converge to 
approximately the same value as $L$ increases.  
This convergence point can be identified as the QCP. Note that whether a given $\Delta_n$ exhibits a zero crossing or a local minimum also depends on the value of $h^z$.


\begin{figure}[t]
    \centering
    \includegraphics[width=11cm]{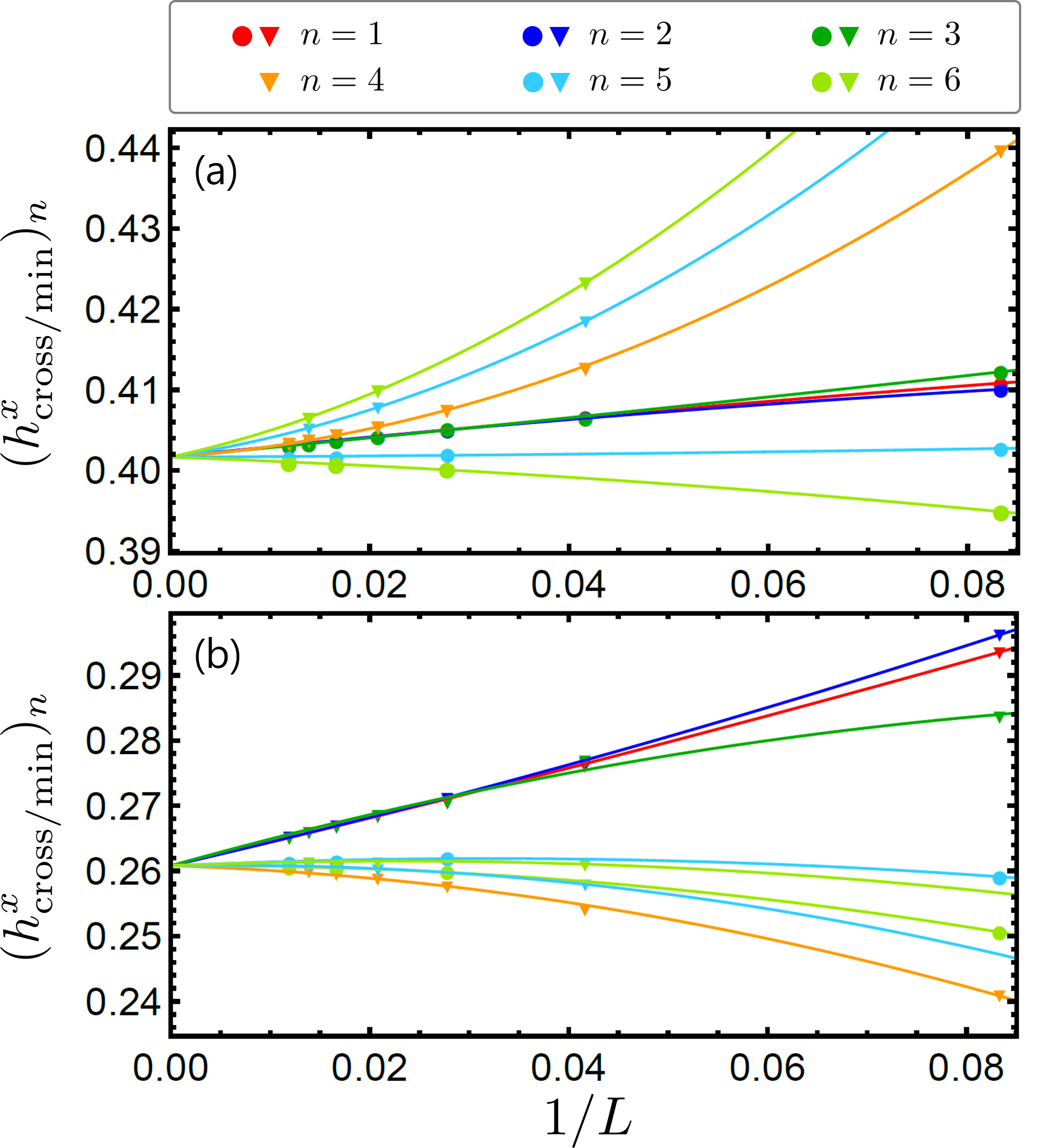}
    \caption{
    $(h^x_{\mathrm{cross}})_n$ (circles) and $(h^x_{\mathrm{min}})_n$ (triangles) for the ground state of the nearest-neighbor model with SSD [Eq.~(\ref{H_TIM})], 
    shown as functions of $1/L$ at (a) $h^z = 0.5$ and (b) $h^z = 0.75$.  
    The color code is as follows: red ($n = 1$), blue ($n = 2$), green ($n = 3$), orange ($n= 4$), light blue ($n = 5$), and light green ($n= 6$).  
    Each solid line represents a least-squares fit (see Appendix~\ref{appendix_fitting} for the fitting parameters).
    }
    \label{TIM_scaling}
\end{figure}

Figures~\ref{TIM_scaling}(a) and~\ref{TIM_scaling}(b) show the size dependence of $(h^x_{\mathrm{cross}})_n$ and $(h^x_{\mathrm{min}})_n$ plotted against $1/L$ for $h^z = 0.5$ and $0.75$, respectively.  
In Fig.~\ref{TIM_deltas}, results for $h^z = 0.5$ up to $L = 48$ are displayed for clarity, while larger system sizes up to $L = 84$ ($L = 60, 72,$ and $84$) were also included in the extrapolation analysis.  
As shown in Figs.~\ref{TIM_scaling}(a) and~\ref{TIM_scaling}(b), $(h^x_{\mathrm{cross}})_n$ and $(h^x_{\mathrm{min}})_n$ form distinct series even for the same $n$, 
while all data points for different $\Delta_n$ and both types of quantities converge toward a single value as $L$ increases, indicating that they share the same thermodynamic limit. 
To estimate the QCP quantitatively, we performed a least-squares fit using the polynomial form  
\begin{equation}
    (h^x_{\mathrm{cross/min}})_n = p_n^{(2)}\!\left(\frac{1}{L}\right)^2 + p_n^{(1)}\!\left(\frac{1}{L}\right) + p^{(0)},\label{fittings}
\end{equation}
where $p^{(0)}$ is taken to be common for all $n$ and both types (cross and min) so that the QCP value of $h^x$ can be determined more precisely.  
The fitted coefficients $p_n^{(1)}$ and $p_n^{(2)}$ are listed in Appendix~\ref{appendix_fitting}.
The extrapolated quantum critical points in the thermodynamic limit are obtained as $h^x_{\mathrm{c}} = 0.40165(7)$ for $h^z = 0.5$ and $h^x_{\mathrm{c}} = 0.26080(28)$ for $h^z = 0.75$. 
Remarkably, even when the fitting is restricted to data up to $L = 36$, the estimated values remain consistent within the statistical error, $h^x_{\mathrm{c}} = 0.402(4)$ and $h^x_{\mathrm{c}} = 0.260(1)$, demonstrating that our SSD-based approach enables highly accurate determination of the QCP from relatively small system sizes.
 
This high accuracy can be attributed to the nature of our criterion, which is based on the condition that local expectation values become equal across all sites.  
Because many independent combinations of local observables can be constructed from this condition, multiple sequences of $(h^x_{\mathrm{cross}})_n$ and $(h^x_{\mathrm{min}})_n$ are available for extrapolation.  
These sequences include trends converging from both above and below toward the critical point, which collectively lead to a highly robust convergence to a single value of $h^x_{\mathrm{c}}$.

\begin{figure}[t]
    \centering
    \includegraphics[width=10cm]{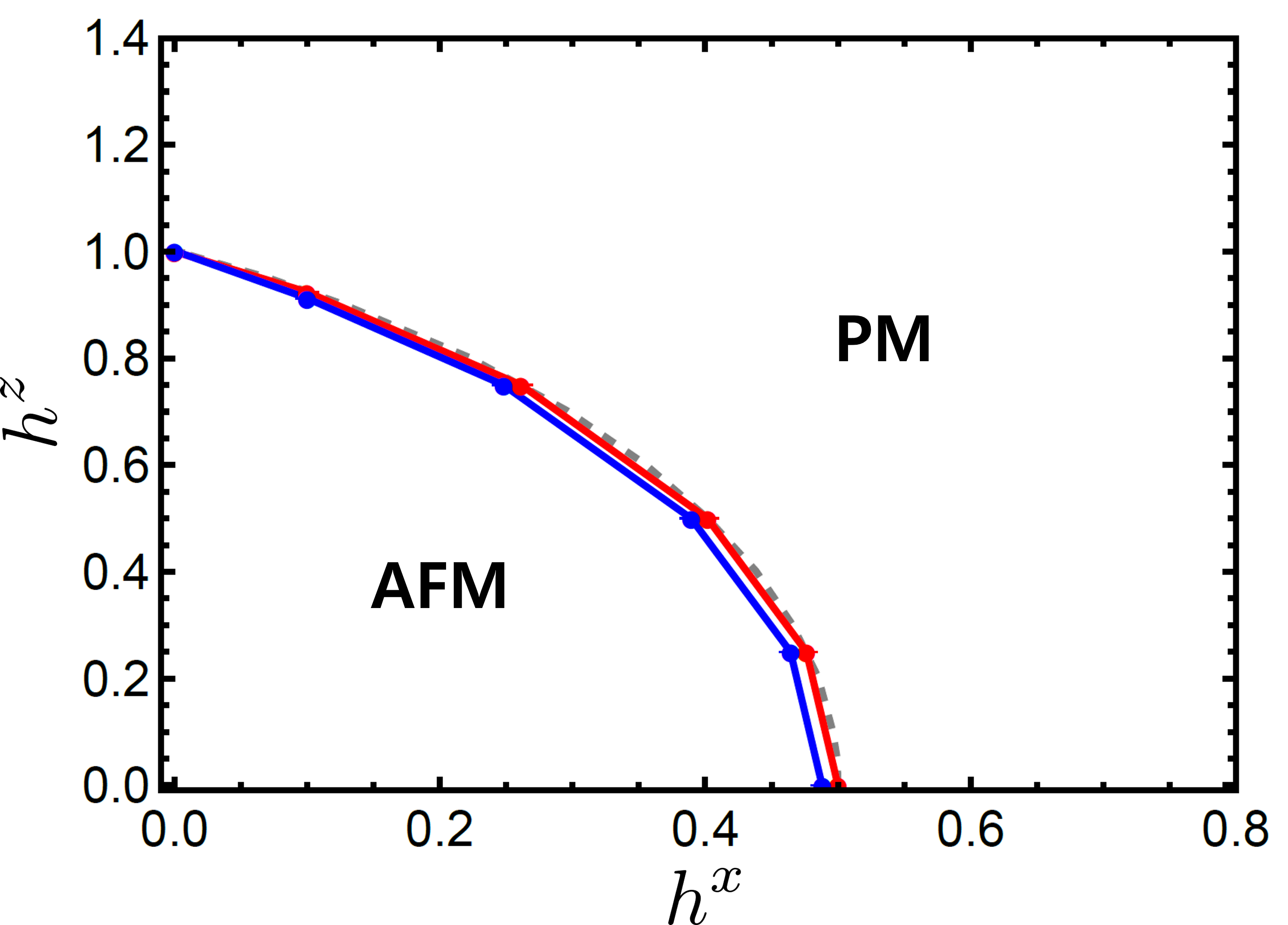}
    \caption{Ground-state phase diagram of the mixed-field Ising chain in the $h^x$–$h^z$ plane.  
    Red and blue points represent the QCPs of the mixed-field Ising chain with nearest-neighbor and long-range interactions, respectively, determined by our SSD-based analysis using system sizes up to $L = 84$.  
    The gray dashed line indicates the phase boundary of the nearest-neighbor model extracted from Fig. 1 in Ref.~\cite{Ising_pd}.
    }
   \label{phasediagram}
\end{figure} 

We performed the same analysis for several other values of $h^z$, and the extrapolated QCPs obtained from system sizes up to $L = 84$ are summarized in the phase diagram shown in Fig.~\ref{phasediagram}.  
Each point represents the critical value of $h^x_{\mathrm{c}}$ determined from the finite-size scaling of $(h^x_{\mathrm{cross}})_n$ and $(h^x_{\mathrm{min}})_n$, while the solid line connecting the data points serves as a guide to the eye. 
For comparison, we also show as a dashed line the phase boundary reported in Ref. \cite{Ising_pd}, where the QCP was identified from the closing of the excitation gap using the infinite-size DMRG algorithm \cite{DMRG1,DMRG2}. Despite using only system sizes up to $L\leq84$, the SSD-based analysis yields results in very good agreement with those of Ref. \cite{Ising_pd}.

As a quantitative comparison, we perform finite-size DMRG calculations under PBCs for $L\leq300$ at $h^z=0.5$ to obtain the excitation gaps, and estimate the QCP $h^x_{\rm c}=0.40205(7)$ using the gap-closing criterion (see Appendix \ref{appendix_comparison_NN}). 
The SSD-based estimate $h^x_{\rm c}=0.40165(7)$, obtained using only system sizes up to $L\leq84$ [Fig. \ref{TIM_scaling}(a)], shows very good agreement with this value, with a relative error of about $0.1\%$, despite the significantly smaller system sizes. 
In contrast, when the same system-size range ($L\leq84$) is used within the gap-closing approach, the QCP cannot be reliably determined, because the excitation gap varies smoothly near the critical point, making it difficult to identify the gap-closing point. 
This clearly demonstrates the advantage of the SSD-based approach in mitigating finite-size effects.
Moreover, from an experimental perspective, the gap-closing criterion generally requires access to excitation spectra, for example via spectroscopic probes such as photon absorption techniques. 
In contrast, the SSD-based method offers a clear advantage as it relies only on local observables that can be measured relatively easily, given that current artificial quantum platforms allow direct access to individual atoms or qubits (see also Sec. \ref{sec3}).

\subsection{Long-range interaction model}

We now turn to the mixed-field antiferromagnetic spin-$1/2$ Ising chain with long-range interactions, $\hat{\mathcal{H}}_{\mathrm{LR}}$ [Eq.~\eqref{H_TIM_LR_bare}].  
This model includes additional power-law decaying couplings $\propto 1/|i-j|^{6}$, which correspond to the typical form of the van der Waals interaction~\cite{Rydberg2,Rydberg5,Rydberg6}.  
We apply the same SSD-based analysis to the long-range interaction model, whose SSD-deformed Hamiltonian is given by
\begin{equation}
\hat{\mathcal{H}}_{\mathrm{LR}}^{\mathrm{(SSD)}} =
\sum_{i< j}\frac{1}{|i-j|^{6}} f_L\!\left(\tfrac{i+j}{2}\right)\hat{S}_i^z\hat{S}_{j}^z
-h^x\sum_{i=1}^L f_L(i)\hat{S}_i^x
-h^z\sum_{i=1}^L f_L(i)\hat{S}_i^z.
\label{H_TIM_LR}
\end{equation}
As in the nearest-neighbor case, we use the DMRG method to obtain the ground state of $\hat{\mathcal{H}}_{\mathrm{LR}}^{\mathrm{(SSD)}}$ and analyze the site dependence of local observables $\hat{O}_i = \hat{S}_i^x$ to determine the QCP.

\begin{figure}[t]
    \centering
    \includegraphics[width=11cm]{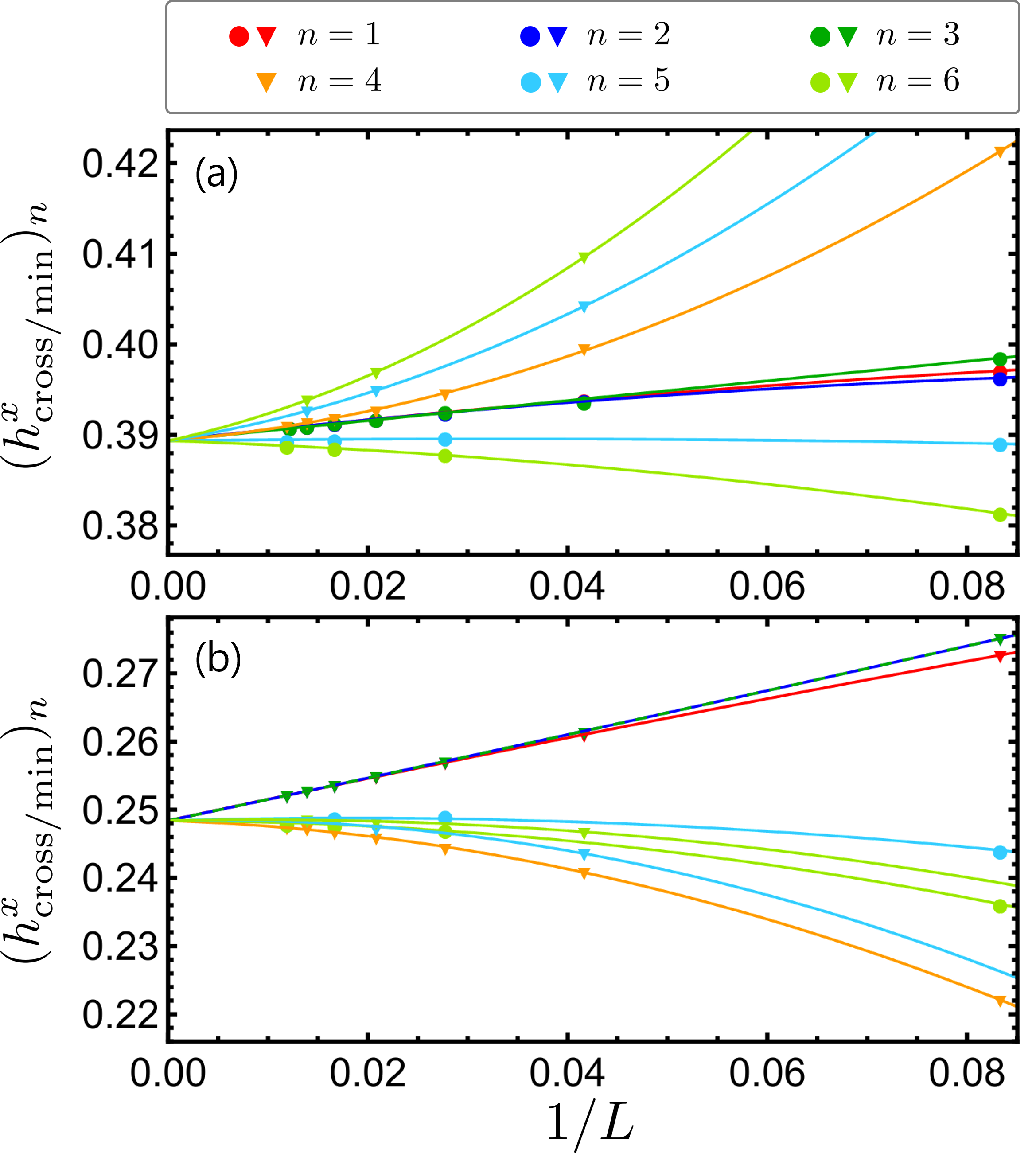}
    \caption{
    Same as Figs.~\ref{TIM_scaling}(a) and~\ref{TIM_scaling}(b), but for the long-range interaction model with SSD [Eq.~\eqref{H_TIM_LR}] at (a) $h^z = 0.5$ and (b) $h^z = 0.75$ (see Appendix~\ref{appendix_fitting} for the fitting parameters).
    }
    \label{TIM_scaling_LR}
\end{figure}

Figures~\ref{TIM_scaling_LR}(a) and~\ref{TIM_scaling_LR}(b) show the size dependence of $(h^x_{\mathrm{cross}})_n$ and $(h^x_{\mathrm{min}})_n$ plotted against $1/L$ for $h^z = 0.5$ and $0.75$, respectively.  
The extrapolated QCPs were obtained from system sizes up to $L = 84$, following the same polynomial fitting procedure as in the nearest-neighbor case.  
The resulting critical points are plotted as blue symbols in the phase diagram of Fig.~\ref{phasediagram}.  
The long-range interaction only slightly modifies the phase boundary: in the classical limit ($h^x = 0$), the transition occurs at $(h^x_{\mathrm{c}}, h^z_{\mathrm{c}}) = (0,1.00145)$, nearly identical to the nearest-neighbor result.  
For smaller $h^z$, however, the critical line shifts slightly toward lower $h^x$ values, reflecting frustration induced by the competing long-range antiferromagnetic couplings.
Our extrapolation using system sizes up to $L = 84$ gives the QCP at $h^z = 0$ as $h^x_{\mathrm{c}} = 0.488019(2)$, which is about $2.4\%$ smaller than that of the nearest-neighbor model. 
This result is in good agreement with the QCP calculated by the gap-closing criterion \cite{Ising_pd} using $L\leq300$ with a relative error of about $0.08\%$ (see Appendix \ref{appendix_comparison_LR}).

\subsection{Scaling analysis of SSD observables}

In the previous subsections, we demonstrated that the SSD-based approach enables precise determination of the QCPs in both nearest-neighbor and long-range quantum Ising chains, even for relatively small system sizes.  
Here, we further examine whether this framework can be extended beyond locating the QCPs to characterize critical phenomena, in particular to extract critical exponents from the scaling behavior of the quantities $\Delta_n$, which represent differences between local observables $\langle \hat S_i^x \rangle$ and $\langle \hat S_j^x \rangle$ under SSD.

\begin{figure}[t]
    \centering
    \includegraphics[width=13cm]{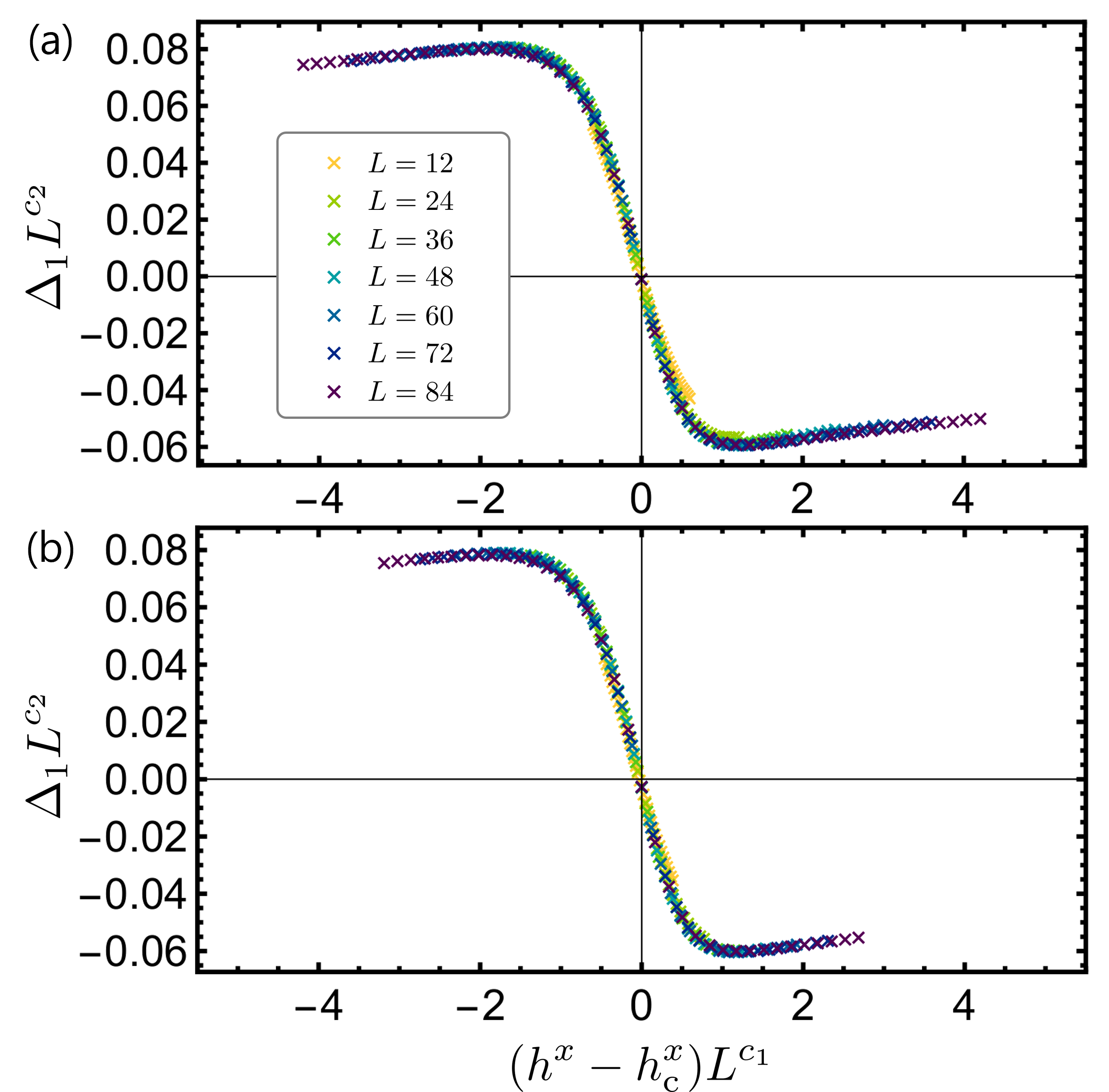}
    \caption{Finite-size scaling collapse of $\Delta_1 L^{c_2}$ versus $(h^x - h^x_{\mathrm{c}}) L^{c_1}$ for (a) nearest-neighbor and (b) long-range interactions at zero longitudinal field ($h^z = 0$). The scaling exponents used are $(c_1, c_2) = (1, 1/8)$.}
   \label{FSS}
\end{figure} 

\begin{figure}[t]
    \centering
    \includegraphics[width=13cm]{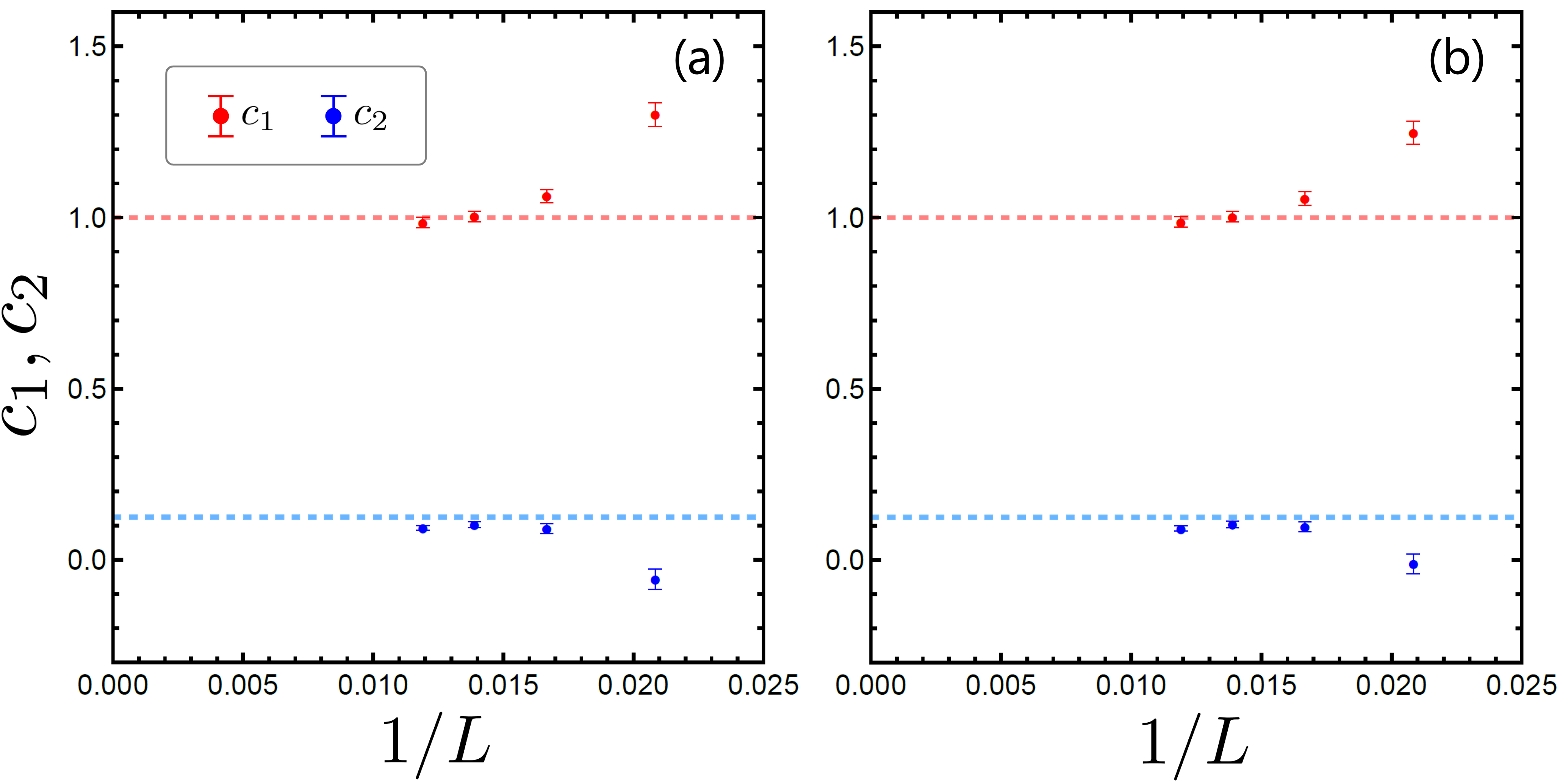}
    \caption{Optimized scaling exponents in Eq.~\eqref{eq:empirical_scaling} obtained from the Bayesian scaling analysis~\cite{FSS1,FSS2}, plotted as functions of $1/L$ for (a) the nearest-neighbor and (b) the long-range interaction models at zero longitudinal field. The red and blue dashed lines indicate the reference values $c_1 = 1$ and $c_2 = 1/8$, respectively.}
   \label{c1c2}
\end{figure}

Although the quantity $\Delta_n$ originates from differences between local observables, it is not \textit{a priori} clear whether it follows the same scaling law as conventional order parameters near the QCP.  
Moreover, $\Delta_n$ is defined within the SSD-deformed Hamiltonian rather than the original uniform model, and it is distinct from the usual order parameters such as the staggered magnetization $\sum_i(-1)^i \hat{S}_i^z $.  
Therefore, there is no theoretical guarantee that $\Delta_n$ should exhibit critical scaling behavior.  
In the following, we focus on $\Delta_1$, which measures the difference between the local expectation values at the center and the edge sites, and examine whether it obeys a universal finite-size scaling form.  
Specifically, we tentatively assume the scaling ansatz  
\begin{equation}
    \Delta_1(h^x,L) = L^{-c_2}\, F\!\left[(h^x-h^x_{\mathrm{c}})L^{c_1}\right],
    \label{eq:empirical_scaling}
\end{equation}
where $c_1$ and $c_2$ are fitting exponents and $F[x]$ is a scaling function.  
We then test whether this ansatz consistently describes the numerical data obtained from the SSD ground states.

To verify the scaling hypothesis in Eq.~\eqref{eq:empirical_scaling},  
we perform a data-collapse analysis of $\Delta_1$ for both the nearest-neighbor and long-range interaction models at $h^z = 0$.  
For each model, the critical field $h^x_{\mathrm{c}}$ is fixed to the value obtained in the previous subsections.  
Figures~\ref{FSS}(a) and~\ref{FSS}(b) show the plots of $\Delta_1 L^{c_2}$ as a function of $(h^x - h^x_{\mathrm{c}}) L^{c_1}$  
for the two models with the choice of scaling exponents $c_1 = 1$ and $c_2 = 1/8$.  
As seen in the figures, the data for different system sizes collapse remarkably well onto a single curve,  
indicating that the critical behavior of $\Delta_1$ is captured by the scaling ansatz of Eq.~\eqref{eq:empirical_scaling} with exponents consistent with the correlation-length exponent $\nu = 1$ and the order-parameter exponent $\beta = 1/8$ of the $(1+1)$-dimensional Ising universality class. 
It is worth noting that $\Delta_1$ is not a conventional symmetry-breaking order parameter,  
so there is no  \textit{a priori} reason for it to follow the Ising value of $\beta$.  
Interestingly, both models exhibit almost the same scaling behavior,  
suggesting that $\Delta_1$ may inherit the same universality as the conventional order parameter.

As a more refined analysis, we further estimate the exponents $c_1$ and $c_2$ using the Bayesian scaling analysis (BSA) method~\cite{FSS1,FSS2}.  
To examine how the extracted exponents depend on the accessible system size,  
we apply the BSA to sliding windows consisting of four consecutive sizes of the form  
$\{L-36,\, L-24,\, L-12,\, L\}$, where $L$ denotes the largest system size in the window.  
For each window, the BSA yields an optimal pair of exponents $(c_1, c_2)$ that best collapse the data within that size range.  
Figures~\ref{c1c2}(a) and~\ref{c1c2}(b) show the resulting values of $c_1$ and $c_2$ plotted as functions of $1/L$.  
While the convergence is not entirely systematic within the present range of system sizes,  
both exponents exhibit a gradual trend toward the expected Ising values as $1/L$ decreases,  
suggesting that $\Delta_1$ indeed carries information on the underlying critical behavior.

Although a more systematic verification is required,  
these results indicate that the SSD-based framework may provide a useful approach for extracting critical exponents  
and analyzing scaling behavior in other quantum systems.  
An important advantage is that the quantities $\Delta_n$ are defined solely from local expectation values in the SSD ground state  
and therefore remain well defined even in systems without a conventional local order parameter.  
This raises the intriguing possibility that similar scaling analyses could be applied to phases with topological or symmetry-protected order,  
where standard order parameters are absent.  
Furthermore, although we focused mainly on $\Delta_1$,  
all $\Delta_n$ can in principle be used simultaneously,  
potentially offering multiple independent scaling conditions that improve the precision of exponent estimation.  
Exploring these directions, including larger system sizes and the combined use of several $\Delta_n$,  
constitutes an interesting open problem for future work.

\section{Implementation in Rydberg quantum simulators} \label{sec3}
The SSD-based approach developed in this work relies only on the ground state of a finite open system, which makes it particularly suitable for current quantum simulators and NISQ processors, where the accessible system sizes are typically limited and periodic boundary conditions are difficult to realize~\cite{NISQ1,NISQ2,NISQ3}.  
In gate-based quantum processors, the SSD Hamiltonian can, in principle, be implemented by appropriately modulating the local gate amplitudes according to the spatial envelope $f_L(i)$.  
Such an implementation may be useful, for example, for efficient ground-state preparation of critical systems that possess translational symmetry.  
For analog quantum simulators, the implementation method depends on the specific platform. 
Here, we take Rydberg atom arrays~\cite{Rydberg1,Rydberg2,Rydberg3,Rydberg4,Rydberg5,Rydberg6} as a representative platform, where long-range van der Waals interactions proportional to $1/r^6$ arise naturally and the atomic geometry can be flexibly programmed using optical tweezers.

We now discuss how SSD Hamiltonians can be implemented in Rydberg quantum simulators. Here, we consider spin models with both nearest- and next-nearest-neighbor antiferromagnetic interactions, that is, the so-called antiferromagnetic $J_1$–$J_2$ interactions~\cite{J1J2}, can be realized within such a Rydberg-atom array.  
This type of interaction pattern is particularly interesting because it introduces frustration and competition between couplings at different distances, which are key ingredients in many strongly correlated quantum systems.  
Note that the $J_1$–$J_2$ interaction considered here is merely an illustrative example, and similar implementations of SSD Hamiltonians could in principle be extended to other types of couplings, such as the $XXZ$ coupling \cite{Ryd_XXZ1,Ryd_XXZ2,Ryd_XXZ3,Ryd_XXZ4,Ryd_XXZ5,Ryd_XXZ6,Ryd_XXZ7,Ryd_XXZ8}.
In the present context, it serves to demonstrate how the spatially dependent interactions required by the SSD Hamiltonian can be engineered by arranging Rydberg atoms with controllable separations in optical tweezers.
As a byproduct, we will show how an appropriate geometric configuration allows a very good approximation of the nearest-neighbor model of the previous sections.

The target Hamiltonian to be implemented is the SSD version of the $J_1$–$J_2$ Ising chain, given by
\begin{equation}
\hat{\mathcal{H}}_{J_1\textup{-}J_2}^{\rm (SSD)}=J_1\sum_{i=1}^{L-1}f_L\left(i+\tfrac{1}{2}\right)\hat{S}_i^z\hat{S}_{i+1}^z+J_2\sum_{i=1}^{L-2}f_L\left(i+1\right)\hat{S}_i^z\hat{S}_{i+2}^z-h^x\sum_{i=1}^L f_L(i)\hat{S}_i^x
-h^z\sum_{i=1}^L f_L(i)\hat{S}_i^z,
\label{H_J1J2_SSD}
\end{equation}
where $J_1>0$, $J_2>0$ are exchange couplings between the nearest-neighbor and the next-nearest-neighbor atoms, respectively.

\begin{figure}[t]
    \centering
    \includegraphics[width=7cm]{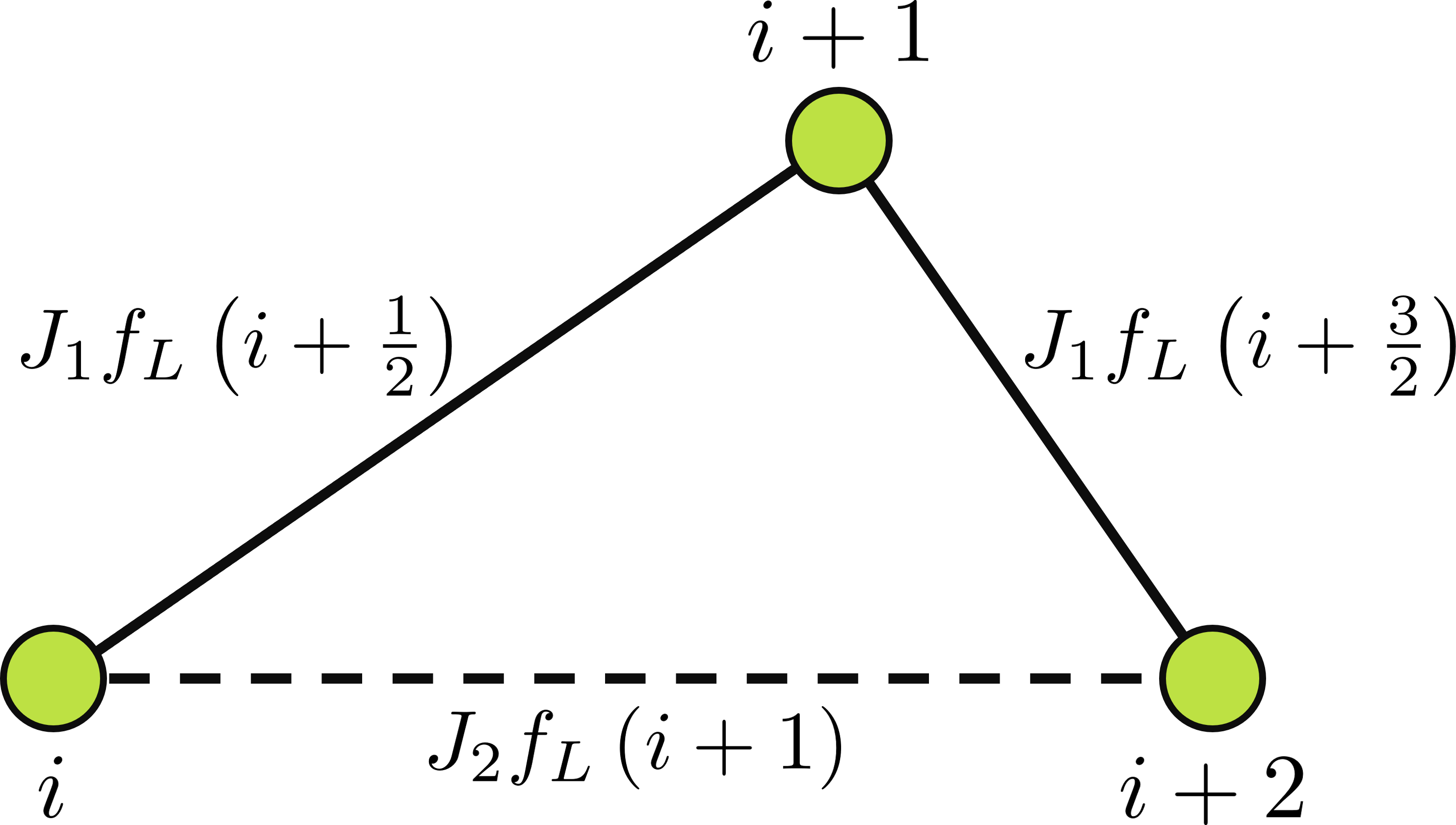}
    \caption{Schematic illustration of the three-atom arrangement corresponding to the SSD-modulated $J_1$–$J_2$ Ising couplings between sites $i$, $i\!+\!1$, and $i\!+\!2$.}
   \label{imp_tri}
\end{figure}

In a Rydberg-atom array, the effective Ising-type coupling between two atoms located at positions $\bm{r}_i$ and $\bm{r}_j$ arises from the van der Waals interaction,
\begin{equation}
V_{ij} = \frac{C_6}{|\bm{r}_i - \bm{r}_j|^6},
\label{eq:vdW}
\end{equation}
where $C_6$ is the state-dependent van der Waals coefficient.  
By arranging the atoms at positions $\{\bm{r}_i\}$ with controllable separations, one can design the spatial profile of the couplings $V_{ij}$ between the $i$-th and $j$-th sites.  
By focusing on three consecutive sites $(i,\, i\!+\!1,\, i\!+\!2)$ [see Fig.~\ref{imp_tri}], one obtains three local coupling conditions required to reproduce the SSD Hamiltonian~\eqref{H_J1J2_SSD}, corresponding to the distances between these atoms:
\begin{eqnarray}
\frac{C_6}{|\bm{r}_{i+1}-\bm{r}_i|^6}&=&J_1 f_L\!\left(i+\tfrac{1}{2}\right), \label{eq:J1_local}\\
\frac{C_6}{|\bm{r}_{i+2}-\bm{r}_{i+1}|^6}&=&J_1 f_L\!\left(i+\tfrac{3}{2}\right), \label{eq:J1_local2}\\
\frac{C_6}{|\bm{r}_{i+2}-\bm{r}_i|^6}&=&J_2 f_L(i+1). \label{eq:J2_local}
\end{eqnarray}
From these relations, the atomic positions can be recursively determined using
Eqs.~(\ref{imp_SSD_dist1}) and~(\ref{imp_SSD_dist2}) below,
which define the successive distance ratios required to realize the SSD-modulated $J_1$–$J_2$ interactions:
\begin{equation}
    \frac{|\bm{r}_{i+2}-\bm{r}_{i+1}|}{|\bm{r}_{i+1}-\bm{r}_i|}=
   \left[\frac{\sin\left(\frac{\pi}{L}i\right)}{\sin\left(\frac{\pi}{L}(i+1)\right)}\right]^{\frac{1}{3}}
\label{imp_SSD_dist1}
\end{equation}
and
\begin{equation}
  \frac{|\bm{r}_{i+2}-\bm{r}_{i}|}{|\bm{r}_{i+1}-\bm{r}_i|} =
\left[\frac{J_1\sin^2\left(\frac{\pi}{L}i\right)}{J_2\sin^2\left(\frac{\pi}{L}\left(i+\frac{1}{2}\right)\right)}\right]^{\frac{1}{6}}
\label{imp_SSD_dist2}
\end{equation}
for $i = 1, 2, \ldots, L-2$.

It should be noted that the interatomic distances must satisfy the triangle inequality, namely, the next-nearest-neighbor bond length cannot exceed the sum of the two nearest-neighbor bond lengths connecting them:
\begin{eqnarray}
|\bm{r}_{i+1}-\bm{r}_i| + |\bm{r}_{i+2}-\bm{r}_{i+1}| > |\bm{r}_{i+2}-\bm{r}_i|.
\label{eq:triangle_ineq}
\end{eqnarray}
This imposes a geometric constraint on the realizable coupling ratios $J_2/J_1$:
\begin{equation}
\left(\frac{J_2}{J_1}\right)_{\!\min}
< \frac{J_2}{J_1},
\label{eq:J2J1_range}
\end{equation}
where 
\begin{eqnarray}
   \left(\frac{J_2}{J_1}\right)_{\!\min}&=&\max_i\left(\left[\frac{
   \sqrt[3]{\frac{\sin\left(\frac{\pi}{L}i\right)}{\sin\left(\frac{\pi}{L}\left(i+\frac{1}{2}\right)\right)}}
    }{ 1+\sqrt[3]{\frac{\sin\left(\frac{\pi}{L}i\right)}{\sin\left(\frac{\pi}{L}\left(i+1\right)\right)}}}\right]^{6}\right)\nonumber\\
    &=& \frac{1}{64} {\cos^2\left(\frac{\pi}{2L}\right)},
   \quad \text{at } i = \frac{L-1}{2}.
\end{eqnarray}
The values of $\left(J_2/J_1\right)_{\rm \min}$ almost saturate to $1/64\approx 0.0156$ for $L\gtrsim 100$. 
Hence, although the case of $J_2/J_1 = 0$ cannot be realized strictly, the ratio $J_2/J_1$ can in practice be reduced to below 2\% by choosing an appropriate geometry, indicating that the nearest-neighbor Ising chain with SSD [Eq.~\eqref{H_TIM}] can be realized to a very good approximation in realistic Rydberg-atom experiments.

\begin{figure*}[h]
    \centering
    \includegraphics[width=14cm]{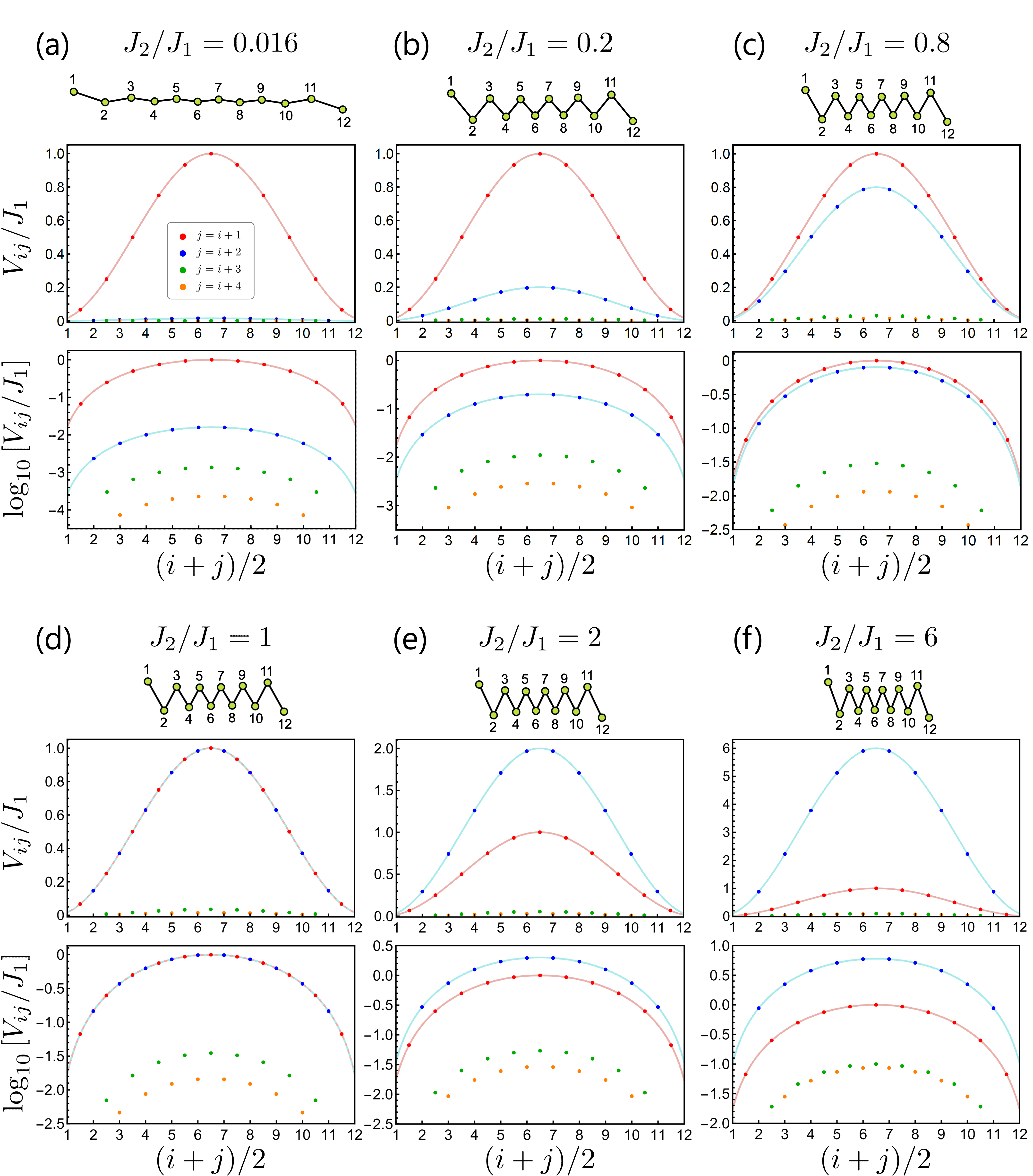}
    \caption{
Configurations of Rydberg atoms and the corresponding spatial profiles of the van der Waals interactions $V_{ij}$ 
for implementing the 12-site $J_1$–$J_2$ Ising chain with SSD at $J_2/J_1 =$ (a) 0.016, (b) 0.2, (c) 0.8, (d) 1, (e) 2, and (f) 6, respectively.  
The upper panels show the zigzag atomic configurations, determined up to an overall length scale.  
The middle panels display $V_{ij}/J_1$ as functions of the midpoint position $(i + j)/2$ for $n$-th-neighbor pairs $(i,j= i + n)$ with $n = 1, 2, 3,$ and $4$.  
The red and blue solid lines represent the functions $f_L(i)$ and $(J_2/J_1)f_L(i)$, respectively.  
The bottom panels show the base-10 logarithm of $V_{ij}/J_1$, corresponding to the middle panels.
}
   \label{imp_SSD}
\end{figure*}

Using the geometric relations Eqs.~\eqref{imp_SSD_dist1} and~\eqref{imp_SSD_dist2}, 
we determined the atomic positions sequentially to realize the desired ratios of $J_2/J_1$ for $L=12$.  
The resulting zigzag configurations, defined up to an overall length scale, and the corresponding spatial profiles of the van der Waals interactions $V_{ij}$ 
are shown in Figs.~\ref{imp_SSD}(a–f) for several representative values of $J_2/J_1 = 0.016,\, 0.2,\, 0.8,\, 1,\, 2,$ and $6$.  
As seen in the figures, the intended antiferromagnetic $J_1$–$J_2$ interactions with SSD are well reproduced in the constructed geometries.  
The geometric conditions in Eqs.~\eqref{imp_SSD_dist1} and~\eqref{imp_SSD_dist2} constrain only the ratio $J_2/J_1$, 
and thus do not explicitly fix longer-range couplings.  
Nevertheless, the plots of $V_{ij}$ show that third- and higher-neighbor interactions are always much weaker than the nearest- and next-nearest-neighbor ones.  
For example, at $J_2/J_1 = 0.016$, which is close to the lower bound of the realizable ratio, 
the third-neighbor coupling is about three orders of magnitude smaller than the nearest-neighbor one, 
showing that the SSD-deformed nearest-neighbor Ising coupling in Eq.~\eqref{H_TIM} is realized to a very good approximation.  
For moderate ratios $J_2/J_1 \lesssim 1$, covering the physically most interesting regime, 
the higher-order interactions remain about two orders smaller, confirming that the target $J_1$–$J_2$ couplings are well realized.  
When $J_2/J_1$ is increased further, the third- and higher-neighbor terms gradually become more relevant; 
around $J_2/J_1 \sim 6$, their magnitudes reach roughly 10\% of the nearest-neighbor coupling, producing non-negligible effects.

The SSD-deformed magnetic-field terms in Eq.~\eqref{H_J1J2_SSD} can in principle be realized through site-dependent Rabi frequencies and detunings, respectively.  
In Rydberg-atom platforms, such spatial control can be achieved using acousto-optic deflectors or spatial light modulators,  
which enable high-resolution shaping of the intensity and phase of the driving lasers~\cite{AOD1,AOD2,SLM}.  
The resulting position-dependent Rabi frequency $\Omega_i$ and detuning $\Delta_i$, realize effective fields proportional to the SSD envelope $f_L(i)$.  
These techniques, already established in programmable Rydberg-atom experiments~\cite{Rydberg3,Ryd_local1,Ryd_local2}, 
make the realization of both the interaction and field terms of the SSD Hamiltonian feasible with current technology.

\clearpage

\section{Conclusions}
\label{sec4}

We have presented an efficient and scalable method for identifying QCPs in finite quantum systems 
by exploiting the spatial uniformity of local observables in the ground state of Hamiltonians with SSD.  
This approach enables precise determination of QCPs even for relatively small open systems, 
since the SSD effectively suppresses boundary effects and restores translational symmetry at criticality.  
We have demonstrated the validity of this method for both nearest-neighbor and long-range mixed-field Ising chains, 
and shown that the critical points extracted from the finite-size scaling of SSD observables agree excellently with those obtained by conventional methods.  
The same framework also allows a scaling analysis of SSD observables, 
suggesting that critical exponents can be estimated within this formulation.

Although the SSD was originally introduced as a numerical technique to mitigate boundary effects, 
our results indicate that its underlying concept can be physically realized in quantum simulators.  
In particular, we have discussed how the SSD Hamiltonian can be implemented in Rydberg-atom arrays, 
where both the interaction and magnetic-field terms can be engineered through control of atomic geometry and laser parameters.  
This finding bridges the gap between numerical boundary-conditioning methods and experimentally accessible Hamiltonian engineering, 
highlighting the potential of the present approach as a diagnostic tool for quantum criticality 
in programmable quantum simulators and other NISQ devices.

Future directions include applying this method to transitions between gapless phases, 
higher-dimensional systems, and models with topological or nonlocal order, where conventional order parameters are difficult to define. 
In higher-dimensional systems, critical points are not always described by CFTs. 
For example, while the critical points of the 2D transverse-field Ising model and the quantum Heisenberg model are believed to be described by relativistic CFTs, other transitions such as the deconfined quantum critical point \cite{2D_future1} are not described by standard CFTs. 
Since the theoretical basis of our SSD protocol relies on the CFT description of criticality, its applicability to such cases is not immediately clear. 
Whether suitably modified deformation schemes could capture characteristic signatures of criticality in these situations remains an open and interesting direction for future research.
Moreover, even for topological phases characterized by nonlocal order parameters, our approach may still have the potential to detect signatures of criticality. 
Since our criterion relies on the restoration of translational invariance in local observables, the operator used for the diagnosis does not need to be an explicit order parameter. 
For example, the one-dimensional transverse-field Ising model studied here can be interpreted as a $\mathbb{Z}_2 $ symmetry-breaking transition in the spin representation, while in the fermionic representation it corresponds to a topological phase transition. 
This illustrates that local diagnostics can, in some cases, signal transitions that also admit a topological interpretation.
Extending the present approach to more general topological systems remains an important subject for future investigation.

Finally, we note that the exact equivalence between the SSD and PBC ground states has been rigorously proven only for certain exactly solvable one-dimensional models, 
such as free-fermion and transverse-field $XY$ chains~\cite{SSD_CFT1}.  
Whether a similar correspondence holds in more general interacting systems, higher-dimensional models, 
or beyond conformal field theories remains an open question.  
In this regard, the present approach may also serve as a practical diagnostic tool to identify or test the possible emergence of SSD–PBC equivalence 
in regimes where analytical proofs have not been established.

\section*{Acknowledgements}
We thank G.~De~Chiara, M.~Kunimi, L.~Lepori, D.~Rossini, T.~Tomita and E.~Tonni for useful discussions.


\paragraph{Funding information}
This work was supported by JSPS KAKENHI 
Grant Nos. 21H05185 (G.M, D.Y.), 23K25830 (D.Y.), 24K06890 (D.Y.), and JST PRESTO Grant No. JPMJPR245D (D.Y.).

\begin{appendix}
\numberwithin{equation}{section}

\section{List of fitting parameters}
\label{appendix_fitting}
In this Appendix, we summarize the fitting parameters used in the extrapolation analysis of the QCPs shown in Figs. \ref{TIM_scaling} and \ref{TIM_scaling_LR}.  
The data of $(h^x_{\mathrm{cross/min}})_n$ for $n=1$–$6$ were fitted to the polynomial form
of Eq.~\eqref{fittings}, where $p^{(0)}$ is a common fitting parameter representing the extrapolated critical point in the thermodynamic limit.  
The obtained fitting coefficients are listed in Tables~\ref{table1}--\ref{table4}.

\setlength{\tabcolsep}{12pt}
\begin{table*}[h]
    \caption{Fitting parameters for the nearest-neighbor interaction model with $h^z=0.5$.}
        \centering
        \begin{tabular}{ccccc}
       $n$& type & $p^{(2)}_n$ & $p^{(1)}_n$ & $p^{(0)}$\\
       \hline
       1 & cross & $-0.22(6)$ & $0.128(5)$& $0.40165(7)$\\
       2 & cross & $-0.36(6)$ & $0.131(5)$& $0.40165(7)$\\
       3 & cross & $0.11(6)$& $0.118(5)$& $0.40165(7)$\\
       4 & min & $4.41(6)$& $0.088(5)$& $0.40165(7)$\\
       5 & cross & $0.08(7)$& $0.006(7)$& $0.40165(7)$\\
       5 & min & $5.19(25)$& $0.189(10)$& $0.40165(7)$\\
       6 & cross & $-0.44(7)$& $-0.045(7)$& $0.40165(7)$\\
       6 & min & $5.97(25)$& $0.271(10)$& $0.40165(7)$\\
       \hline
    \end{tabular}\label{table1}
\end{table*}

\begin{table*}[h]
    \caption{Fitting parameters for the nearest-neighbor interaction model with $h^z=0.75$.}
        \centering
        \begin{tabular}{ccccc}
       $n$& type & $p^{(2)}_n$ & $p^{(1)}_n$ & $p^{(0)}$\\
       \hline
       1 & min & $0.43(23)$& $0.358(20)$& $0.26080(28)$\\
       2 & min & $0.87(23)$& $0.352(20)$& $0.26080(28)$\\
       3 & min & $-1.78(23)$& $0.426(20)$& $0.26080(28)$\\
       4 & min & $-2.28(23)$& $-0.050(20)$& $0.26080(28)$\\
       5 & cross & $-1.06(28)$& $0.068(25)$& $0.26080(28)$\\
       5 & min & $-2.27(96)$& $0.026(40)$& $0.26080(28)$\\
       6 & cross & $-1.54(28)$& $0.006(25)$& $0.26080(28)$\\
       6 & min & $-1.33(96)$& $0.061(40)$& $0.26080(28)$\\
       \hline
\end{tabular}\label{table2}
\end{table*}

\clearpage

\begin{table*}[t]
    \caption{Fitting parameters for the long-range interaction model with $h^z=0.5$.}
        \centering
        \begin{tabular}{ccccc}
       $n$& type & $p^{(2)}_n$ & $p^{(1)}_n$ & $p^{(0)}$\\
       \hline
       1 & cross & $-0.37(5)$& $0.124(5)$& $0.38934(6)$\\
       2 & cross & $-0.51(5)$& $0.126(5)$& $0.38934(6)$\\
       3 & cross & $-0.03(5)$& $0.112(5)$& $0.38934(6)$\\
       4 & min & $3.48(5)$& $0.094(5)$& $0.38934(6)$\\
       5 & cross & $-0.21(7)$& $0.014(6)$& $0.38934(6)$\\
       5 & min & $4.28(22)$& $0.178(9)$& $0.38934(6)$\\
       6 & cross & $-0.70(7)$& $-0.037(6)$& $0.38934(6)$\\
       6 & min & $5.89(22)$& $0.241(9)$& $0.38934(6)$\\
       \hline
\end{tabular}\label{table3}
\end{table*}

\begin{table*}[t]
    \caption{Fitting parameters for the long-range interaction model with $h^z=0.75$.}
        \centering
        \begin{tabular}{ccccc}
       $n$  & type & $p^{(2)}_n$ & $p^{(1)}_n$ & $p^{(0)}$\\
       \hline
       1 & min & $-0.29(11)$& $0.315(10)$& $0.24833(14)$\\
       2 & min & $0.12(11)$& $0.310(10)$& $0.24833(14)$\\
       3 & min & $0.12(11)$& $0.310(10)$& $0.24833(14)$\\
       4 & min & $-3.19(11)$& $-0.050(10)$& $0.24833(14)$\\
       5 & cross & $-1.12(14)$& $0.041(12)$& $0.24833(14)$\\
       5 & min & $-1.69(14)$& $-0.006(12)$& $0.24833(14)$\\
       6 & cross & $-3.59(47)$& $0.033(20)$& $0.24833(14)$\\
       6 & min & $-1.68(47)$& $0.031(20)$& $0.24833(14)$\\
       \hline
\end{tabular}\label{table4}
\end{table*}

\section{
Estimate of QCPs by gap-closing criterion}

\subsection{The NN model at $h^z=0.5$}
\label{appendix_comparison_NN}

In this Appendix, we present estimates of the QCPs by the gap-closing criterion \cite{Ising_pd}.

Let us consider the NN model under PBCs, whose Hamiltonian is given by adding the term $\hat{S}_L^z\hat{S}_1^z$, which connects the two edges, to Eq. (\ref{H_TIM_bare}).
Here we focus on the case of $h^z = 0.5$.
For fixed $h^x$, we calculate the two lowest excitation gaps $m_1(L)= E_1(L)-E_0(L)$ and $m_2(L)= E_2(L)-E_0(L)$ for the PBC Hamiltonian (without SSD) using the DMRG algorithm.
Here, $E_0(L)$, $E_1(L)$ and $E_2(L)$ denote the ground-state, first-excited-state, and second-excited-state energies at system size $L$, respectively.
To obtain the energy gaps in the thermodynamic limit $\tilde{m}_\mu=\lim_{L\to\infty} m_\mu(L)$ [$\mu =1,2$], we extrapolate the data for $L=84$, $100$, $200$, and $300$, using the exponential fitting function:
\begin{equation}
m_\mu(L)=A_\mu^{(1)}\exp\left(\frac{A_\mu^{(2)}}{L}\right)+A_\mu^{(3)},
\end{equation}
where $A_\mu^{(1)}$, $A_\mu^{(2)}$, and $A_\mu^{(3)}$ are fitting parameters. 
Here, the thermodynamic-limit value is given by $\tilde{m}_\mu=A_\mu^{(1)}+A_\mu^{(3)}$.

\begin{figure}[h]
    \centering
    \includegraphics[width=15cm]{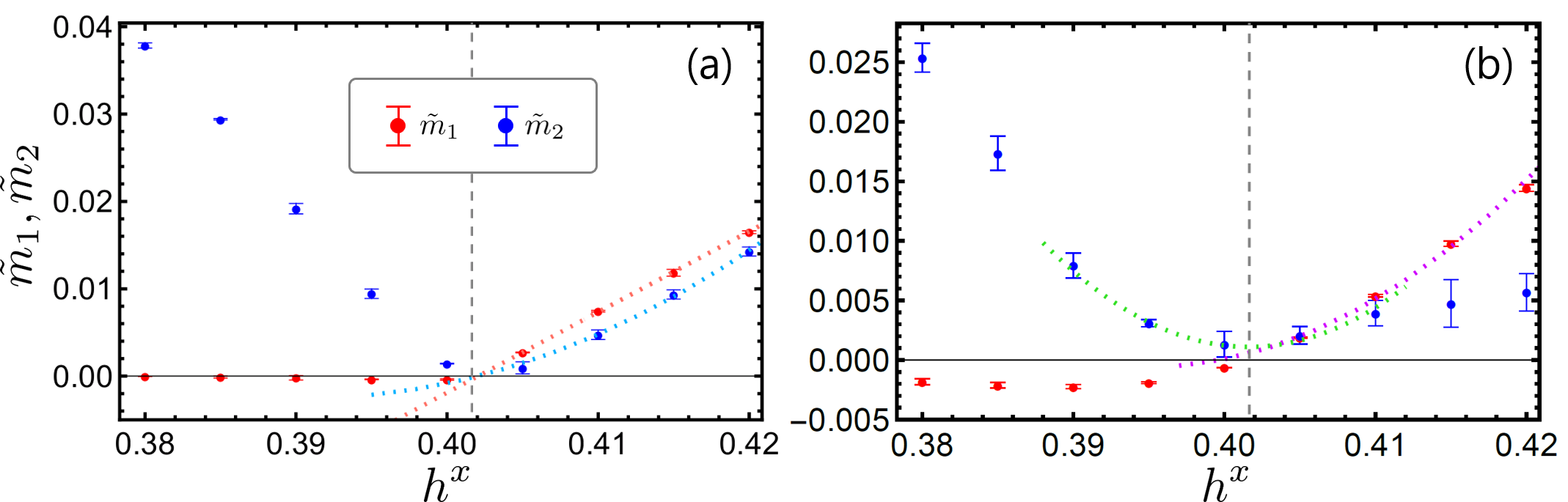}
    \caption{ 
The extrapolated excitation gaps $\tilde{m}_1$ and $\tilde{m}_2$ of the NN model at $h^z=0.5$ obtained from the data for (a) $L=84$, $100$, $200$, $300$ and (b) $L=12$, $24$, $36$, $48$, $60$, $72$, $84$. The red and blue dotted lines in panel (a) and the purple and green dotted lines in panel (b) represent the fits to $\tilde{m}_1$ and $\tilde{m}_2$ (see text for more details). The gray dashed line shows the QCP by the SSD-based analysis with $L\leq 84$.
}
   \label{appendix_energygap_NN}
\end{figure}

By repeating the same procedure for different $h^x$, we obtain $\tilde{m}_\mu$ as  functions of $h^x$ shown in Fig. \ref{appendix_energygap_NN}(a).
In the AFM regime, $\tilde{m}_1$ vanishes because the ground state is doubly degenerate, whereas in the PM regime the ground state is non-degenerate, resulting in a finite $\tilde{m}_1$.
The QCP $h^x_{\rm c}$ is determined from the simultaneous vanishing of $\tilde{m}_1$ and $\tilde{m}_2$.
To estimate the QCP, we further fit the data of $\tilde{m}_\mu$ in the region $h^x\geq 0.405$, where both $\tilde{m}_1$ and $\tilde{m}_2$ take finite positive values in our calculation, using the following linear and quadratic forms:
\begin{eqnarray}
\tilde{m}_1&=&B^{(1)}(h^x-B^{(2)}),
\label{append_fit1}\\
\tilde{m}_2&=&B^{(3)}(h^x-B^{(2)})(h^x-B^{(4)}),
\label{append_fit2}
\end{eqnarray}
where $B^{(1)}$, $B^{(2)}$, $B^{(3)}$, and $B^{(4)}$ are fitting parameters. 
Here $B^{(2)}$ is shared between the two fits so that both fitting functions vanish at the common point $( h^x,\tilde{m}_1)=( h^x,\tilde{m}_2)=(B^{(2)},0)$.
This condition directly implies $ h^x_{\rm c}=B^{(2)}$.
The red and blue dotted lines in Fig. \ref{appendix_energygap_NN}(a) represent these fitting results.
We determine the QCP as $h^x_{\rm c}=0.40205(7)$.
This is consistent with the SSD-based result $h^x_{\mathrm{c}} = 0.40165(7)$, which is shown as the gray dashed line in Fig. \ref{appendix_energygap_NN}, with a relative error of about $0.1\%$.

Restricting the system sizes to $L=12$, $24$, $36$, $48$, $60$, $72$, and $84$, which are the same as those used in our SSD-based analysis, $\tilde{m}_1$ varies smoothly and $\tilde{m}_2$ exhibits a broadened local minimum as shown in Fig. \ref{appendix_energygap_NN}(b), making it difficult to determine the QCP using the fitting forms (\ref{append_fit1}) and (\ref{append_fit2}).
Nevertheless, as a supplementary attempt, we perform simple quadratic fits separately for $\tilde{m}_1$  and $\tilde{m}_2$. 
For $\tilde{m}_1$, we fit the data in the region $h^x\geq 0.405$, yielding $h^x_{\rm c}=0.3997(50)$.
For $\tilde{m}_2$, we fit the data in $0.39\leq h^x\leq 0.41$, and determine the QCP from the minimum point, resulting in $h^x_{\rm c}=0.4016(6)$.
The purple and green dotted lines in Fig. \ref{appendix_energygap_NN}(b) show the quadratic fits.
The relative errors of the former and latter analyses with respect to the result for $L\leq300$ are about $0.58\%$ and  $0.11\%$, respectively.
These errors are larger than that of the SSD-based analysis with $L\leq 84$ [Fig. \ref{TIM_scaling}(a)], which is $0.1\%$.
Although the latter estimate appears quantitatively accurate, we emphasize that the estimate obtained from a shallow and broad minimum may be sensitive to the fitting range and functional form. Therefore, the determination of the QCP based on the gap-closing criterion with small system sizes alone is not robust.


\subsection{The LR model at $h^z=0$}
\label{appendix_comparison_LR}

\begin{figure}[h]
    \centering
    \includegraphics[width=9cm]{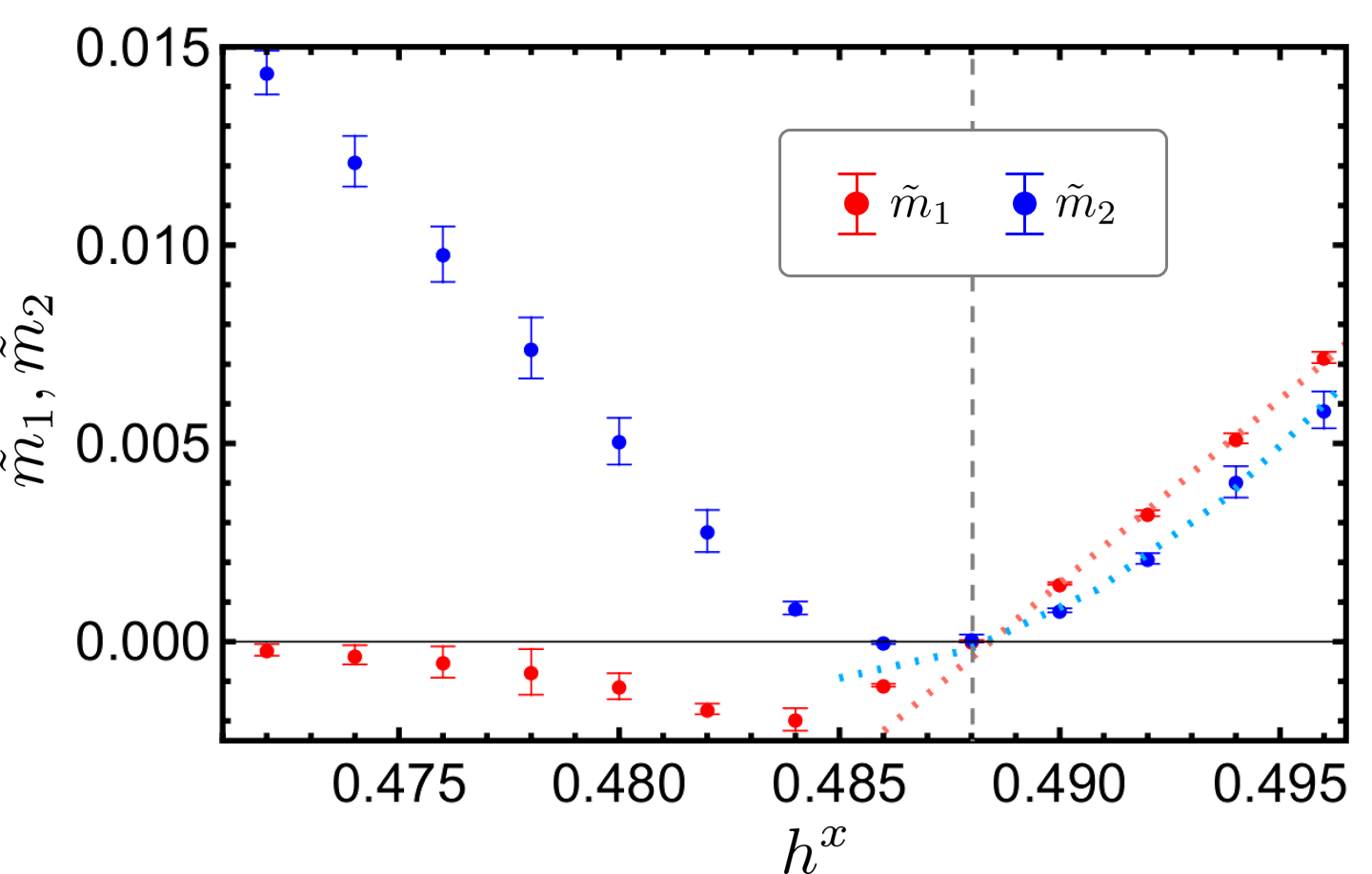}
    \caption{
Same as Fig. \ref{appendix_energygap_NN}(a), but for the LR model at $h^z=0$.
}
   \label{appendix_energygap_LR}
\end{figure}

In this Appendix, as a reference for comparison with the QCP obtained from the SSD-based analysis, we perform the same analysis as in Appendix \ref{appendix_comparison_NN} for the LR model (\ref{H_TIM_LR_bare}) at $h^z=0$.
The extrapolated gaps $\tilde{m}_1$ and $\tilde{m}_2$ obtained from the data for $L=84$, $100$, $200$, and $300$ are shown in Fig. \ref{appendix_energygap_LR}.
The data in the region $h^x\geq0.49$ are fitted using the fitting forms (\ref{append_fit1}) and (\ref{append_fit2}), yielding $h^x_{\rm c}=0.48843(4)$.

\end{appendix}


\clearpage 
\begin{thebibliography}{99}

\bibitem{SSH} Z. Lan, M. L. N. Chen, F. Gao, S. Zhang, and  W. E. I. Sha, {\it A brief review of topological photonics in one, two, and three dimensions}, Reviews in Physics {\bf 9}, 100076 (2022), \doi{
https://doi.org/10.1016/j.revip.2022.100076
}.
\bibitem{topo} M. Z. Hasan and C. L. Kane, {\it Colloquium: Topological insulators}, Reviews of Modern Physics {\bf 82}, 3045 (2010), \doi{https://doi.org/10.1103/RevModPhys.82.3045}.
\bibitem{edge} J. L. Lado, and J. Fern\'{a}ndez-Rossier, {\it Theory of edge states in graphene-like systems}, arXiv:2210.07568 (2022), \doi{
https://doi.org/10.48550/arXiv.2210.07568}.
\bibitem{Haldane1} F. D. M. Haldane, {\it Continuum dynamics of the 1-D Heisenberg antiferromagnet: Identification with the O(3) nonlinear sigma model}, Physics Letters A {\bf93}, 464 (1983), \doi{http://dx.doi.org/10.1016/0375-9601(83)90631-X}.
\bibitem{Haldane2} F. D. M. Haldane, {\it Nonlinear Field Theory of Large-Spin Heisenberg Antiferromagnets: Semiclassically Quantized Solitons of the One-Dimensional Easy-Axis Néel State}, Physical Review Letters {\bf50}, 1153 (1983), \doi{https://doi.org/10.1103/PhysRevLett.50.1153}.

\bibitem{MPS1} G. Vidal, {\it Efficient Classical Simulation of Slightly Entangled Quantum Computations}, Physical Review Letters {\bf 91}, 147902 (2003), \doi{https://doi.org/10.1103/PhysRevLett.91.147902}.
\bibitem{MPS2} S. \"Ostlund and S. Rommer, {\it Thermodynamic Limit of Density Matrix Renormalization}, Physical Review Letters {\bf 75}, 3537 (1995), \doi{https://doi.org/10.1103/PhysRevLett.75.3537}.
\bibitem{PEPS1} F. Verstraete and J. I. Cirac, {\it Renormalization algorithms for Quantum-Many Body Systems in two and higher dimensions}, arXiv:0407066 (2014), \doi{
https://doi.org/10.48550/arXiv.cond-mat/0407066}.
\bibitem{PEPS2} F. Verstraete, M. M. Wolf, D. Perez-Garcia, and J. I. Cirac, {\it Criticality, the Area Law, and the Computational Power of Projected Entangled Pair States}, Physical Review Leters {\bf 96}, 220601 (2006), \doi{https://doi.org/10.1103/PhysRevLett.96.220601}.
\bibitem{PEPS3} V. Murg, F. Verstraete, and J. I. Cirac, {\it Variational study of hard-core bosons in a two-dimensional optical lattice using projected entangled pair states}, Physical Review A {\bf 75}, 033605 (2007), \doi{https://doi.org/10.1103/PhysRevA.75.033605}.

\bibitem{coldatom1} I. Bloch, {\it Ultracold quantum gases in optical lattices}, Nature Physics {\bf 1}, 23 (2005), \doi{https://doi.org/10.1038/nphys138}.
 \bibitem{coldatom2} G. Pagano, M. Mancini, G. Cappellini, P. Lombardi, F. Sch\"{a}fer, H. Hu, X.-J. Liu, J. Catani, C. Sias, M. Inguscio, and L. Fallani, {\it A one-dimensional liquid of fermions with tunable spin}, Nature Physics {\bf 10}, 198 (2014), \doi{https://doi.org/10.1038/nphys2878}.
 \bibitem{coldatom3} S. Nakajima, T. Tomita, S. Taie, T. Ichinose, H. Ozawa, L. Wang, M. Troyer, and Y. Takahashi, {\it Topological Thouless pumping of ultracold fermions}, Nature Physics {\bf 12}, 296 (2016), \doi{https://doi.org/10.1038/nphys3622}.
 \bibitem{trappedion1} R. Blatt and C. F. Roos, {\it Quantum simulations with trapped ions}, Nature Physics {\bf 8}, 277 (2012), \doi{https://doi.org/10.1038/nphys2252}. 
 \bibitem{trappedion2} D. Leibfried, R. Blatt, C. Monroe, and D. Wineland, {\it Quantum dynamics of single trapped ions}, Review of Modern Physics {\bf 75}, 281 (2003), \doi{ https://doi.org/10.1103/RevModPhys.75.281}.
\bibitem{trappedion3} C. Monroe, W. C. Campbell, L. M. Duan, Z. X. Gong, A. V. Gorshkov, P. W. Hess, R. Islam, K. Kim, N. M. Linke, G. Pagano, P. Richerme, C. Senko, and N. Y. Yao, {\it Programmable quantum simulations of spin systems with trapped ions},Reviews of Modern Physics {\bf 93}, 025001 (2021), \doi{https://doi.org/10.1103/RevModPhys.93.025001}.
\bibitem{Rydberg1} F. Nogrette, H. Labuhn, S. Ravets, D. Barredo, L.B\'{e}guin, A.Vernier, T. Lahaye, and A. Browaeys, {\it Single-Atom Trapping in Holographic 2D Arrays of Microtraps with Arbitrary Geometries}, Physical Review X {\bf 4}, 021034 (2014), \doi{https://doi.org/10.1103/PhysRevX.4.021034}. 
 \bibitem{Rydberg2} H. Bernien, S. Schwartz, A. Keesling, H. Levine, A. Omran, H. Pichler, S. Choi, A. S. Zibrov, M. Endres, M. Greiner, V. Vuletić, and M. D. Lukin, {\it Probing many-body dynamics on a 51-atom quantum simulator}, Nature (London) {\bf 551}, 579 (2017), \doi{https://doi.org/10.1038/nature24622}.
 \bibitem{Rydberg3} A. Browaeys and T. Lahaye, {\it Many-body physics with individually controlled Rydberg atoms}, Nature Physics {\bf 16}, 132 (2020), \doi{https://doi.org/10.1038/s41567-019-0733-z}.
 \bibitem{Rydberg4} P. Scholl, M. Schuler, H. J. Williams, A. A. Eberharter, D. Barredo, K.-N. Schymik, V. Lienhard, L.-P. Henry, T. C. Lang, T. Lahaye, A. M. L\"{a}uchli, and A. Browaeys, {\it Quantum simulation of 2D antiferromagnets with hundreds of Rydberg atoms}, Nature (London) {\bf 595}, 233 (2021), \doi{https://doi.org/10.1038/s41586-021-03585-1}.
 \bibitem{Rydberg5} S. Ebadi, T. T. Wang, H. Levine, A. Keesling, G. Semeghini, A. Omran, D. Bluvstein, R. Samajdar, H. Pichler, W. W. Ho, S. Choi, S. Sachdev, M. Greiner, V. Vuleti\'{c}, and M. D. Lukin, {\it Quantum phases of matter on a 256-atom programmable quantum simulator}, Nature (London) {\bf 595}, 227 (2021), \doi{https://doi.org/10.1038/s41586-021-03582-4}.
 \bibitem{Rydberg6} V. Lienhard, S. de L\'{e}s\'{e}leuc, D. Barredo, T. Lahaye, A. Browaeys, M. Schuler, L.-P. Henry, and A.M. L\"{a}uchli, {\it Observing the Space- and Time-Dependent Growth of Correlations in Dynamically Tuned Synthetic Ising Models with Antiferromagnetic Interactions}, Physical Review X {\bf8}, 021070 (2018), \doi{https://doi.org/10.1103/PhysRevX.8.021070}.
 
\bibitem{SBC1} M. Veki\'{c} and S. R. White, {\it Smooth boundary conditions for quantum lattice systems
}, Physical Review Letters {\bf 71}, 4283 (1993), \doi{https://doi.org/10.1103/PhysRevLett.71.4283}.
\bibitem{SBC2} M. Veki\'{c} and S. R. White, {\it Hubbard model with smooth boundary conditions}, Physical Review B {\bf 53}, 14552 (1996), \doi{https://doi.org/10.1103/PhysRevB.53.14552}.
\bibitem{SBC3} T. Hikihara and T. Suzuki, {\it Long-distance entanglement in one-dimensional quantum systems under sinusoidal deformation}, Physical Review A {\bf 87}, 042337 (2013), \doi{https://doi.org/10.1103/PhysRevA.87.042337}.
\bibitem{SBC4} A. Gendiar, M. Dani\u{s}ka, Y. Lee, and T. Nishino, {\it Suppression of finite-size effects in one-dimensional correlated systems}, Physical Review A {\bf 83}, 052118 (2011), \doi{https://doi.org/10.1103/PhysRevA.83.052118}.

\bibitem{SSD1} A. Gendiar, R. Krcmar, and T. Nishino, {\it Spherical Deformation for One-Dimensional Quantum Systems}, Progress of Theoretical Physics {\bf 122}, 953 (2009), \doi{https://doi.org/10.1143/PTP.122.953}; {\it ibid.} {\bf 123}, 393 (2010), \doi{https://doi.org/10.1143/PTP.123.393}.
\bibitem{SSD2} T. Hikihara and T. Nishino, {\it Connecting distant ends of one-dimensional critical systems by a sine-square deformation}, Physical Review B {\bf 83}, 060414(R) (2011), \doi{https://doi.org/10.1103/PhysRevB.83.060414}.
\bibitem{SSD3} H. Katsura, {\it Exact ground state of the sine-square deformed XY spin chain}, Journal of Physics A: Mathematical and Theoretical {\bf 44}, 252001 (2011), \doi{10.1088/1751-8113/44/25/252001}.
\bibitem{SSD4} N. Shibata and C. Hotta, {\it Boundary effects in the density-matrix renormalization group calculation}, Phys. Rev. B {\bf 84}, 115116 (2011), \doi{https://doi.org/10.1103/PhysRevB.84.115116}.
\bibitem{SSD5} I. Maruyama, H. Katsura, and T. Hikihara, {\it Sine-square deformation of free fermion systems in one and higher dimensions}, Physical Review B {\bf 84}, 165132 (2011), \doi{https://doi.org/10.1103/PhysRevB.84.165132}.
\bibitem{SSD6} K. Okunishi and H. Katsura, {\it Sine-square deformation and supersymmetric quantum mechanics}, Journal of Physics A: Mathematical and Theoretical {\bf 48}, 445208 (2015), \doi{10.1088/1751-8113/48/44/445208}.
\bibitem{SSD7} K. Yonaga, and N. Shibata, {\it Ground State Phase Diagram of Twisted Three-Leg Spin Tube in Magnetic Field}, Journal of the Physical Society of Japan {\bf 84}, 094706 (2015), \doi{https://doi.org/10.7566/JPSJ.84.094706}.
\bibitem{c_EE} P. Calabrese and J. Cardy, Journal of Statistical Mechanics: Theory and Experiment, {\it Entanglement entropy and quantum field theory}, P06002 (2004), \doi{10.1088/1742-5468/2004/06/P06002}.
\bibitem{SSD_2D_1} I. Maruyama, H. Katsura, and T. Hikihara, {\it Sine-square deformation of free fermion systems in one and higher dimensions}, Physical Review B {\bf 84}, 165132 (2011), \doi{https://doi.org/10.1103/PhysRevB.84.165132}.
\bibitem{SSD_2D_2} C. Hotta, S. Nishimoto, and N. Shibata, {\it Grand canonical finite size numerical approaches in one and two dimensions: Real space energy renormalization and edge state generation}, Physical Review B {\bf 87}, 115128 (2013), \doi{https://doi.org/10.1103/PhysRevB.87.115128}.
\bibitem{SSD_2D_3} C. Hotta, and K. Asano, {\it Magnetic susceptibility of quantum spin systems calculated by sine square deformation: One-dimensional, square lattice, and kagome lattice Heisenberg antiferromagnets}, Physical Review B {\bf 98}, 140405(R) (2018), \doi{https://doi.org/10.1103/PhysRevB.98.140405}.
\bibitem{SSD_2D_4} S. Nishimoto, N. Shibata, and C. Hotta, {\it Controlling frustrated liquids and solids with an applied field in a kagome Heisenberg antiferromagnet}, Nature communications {\bf 4}, 2287 (2013), \doi{https://doi.org/10.1038/ncomms3287}.
\bibitem{SSD_CFT1} H. Katsura, {\it Sine-square deformation of solvable spin chains and conformal field theories}, Journal of Physics A: Mathematical and Theoretical {\bf 45}, 115003 (2012), \doi{10.1088/1751-8113/45/11/115003}.
\bibitem{SSD_CFT2} T. Tada, {\it Sine-square deformation and its relevance to string theory}, Modern Physics Letters A {\bf 30}, 1550092 (2015), \doi{https://doi.org/10.1142/s0217732315500923}.
\bibitem{SSD_CFT3} N. Ishibashi and T. Tada, {\it Dipolar quantization and the infinite circumference limit of two-dimensional conformal field theories}, International Journal of Modern Physics A {\bf 31}, 1650170 (2015), \doi{https://doi.org/10.1142/S0217751X16501700}.
\bibitem{SSD_CFT4} K. Okunishi, {\it Sine-square deformation and M\"obius quantization of 2D conformal field theory}, Progress of Theoretical and Experimental Physics {\bf 2016}, 063A02 (2016), \doi{https://doi.org/10.1093/ptep/ptw060}.
\bibitem{SSD_CFT5} X. Wen, S. Ryu, and A. W. W. Ludwig, {\it Evolution operators in conformal field theories and conformal mappings: Entanglement Hamiltonian, the sine-square deformation, and others}, Physical Review B {\bf 93}, 235119 (2016), \doi{https://doi.org/10.1103/PhysRevB.93.235119}.
\bibitem{SSD_CFT6} S. Tamura and H. Katsura, {\it Zero-energy states in conformal field theory with sine-square deformation}, Progress of Theoretical and Experimental Physics {\bf 2017}, 113A01 (2017), \doi{https://doi.org/10.1093/ptep/ptx147}.
\bibitem{SSD_CFT7} T. Tada, {\it Conformal quantum mechanics and sine-square deformation}, Progress of Theoretical and Experimental Physics {\bf 2018}, 061B01 (2018), \doi{https://doi.org/10.1093/ptep/pty058}.

\bibitem{NISQ1} J. Preskill, {\it Quantum Computing in the NISQ era and beyond}, Quantum {\bf 2}, 79 (2018), \doi{	https://doi.org/10.22331/q-2018-08-06-79}.
\bibitem{NISQ2} K. Bharti, A. Cervera-Lierta, T. H. Kyaw, T. Haug, S. Alperin-Lea, A. Anand, M. Degroote, H. Heimonen, J. S. Kottmann, T. Menke, W.-K. Mok, S. Sim, L.-C. Kwek, and A. Aspuru-Guzik, {\it Noisy intermediate-scale quantum algorithms}, Review of Modern Physics {\bf 94}, 015004 (2022), \doi{https://doi.org/10.1103/RevModPhys.94.015004}.
\bibitem{NISQ3} J. W. Z. Lau, K. H. Lim, H. Shrotriya, and L. C. Kwek, {\it NISQ computing: where are we and where do we go?}, 
AAPPS bulletin {\bf 32}, 27 (2022), \doi{https://doi.org/10.1007/s43673-022-00058-z}.

\bibitem{DMRG1} S. R. White, {\it Density matrix formulation for quantum renormalization groups}, Physical Review Letters {\bf 69}, 2863 (1992), \doi{https://doi.org/10.1103/PhysRevLett.69.2863}.
\bibitem{DMRG2} S. R. White, {\it Density-matrix algorithms for quantum renormalization groups}, Physical Review B {\bf 48}, 10345 (1993), \doi{https://doi.org/10.1103/PhysRevB.48.10345}.
\bibitem{DMRG3} U. Schollw\"{o}ck, {\it The density-matrix renormalization group in the age of matrix product states}, Annals of physics {\bf 326} 96 (2011), \doi{https://doi.org/10.1016/j.aop.2010.09.012}.

\bibitem{Ising_pd} A. A. Ovchinnikov, D. V. Dmitriev, V. Y. Krivnov, and V. O. Cheranovskii, {\it Antiferromagnetic Ising chain in a mixed transverse and longitudinal magnetic field}, Physical Review B {\bf 68}, 214406 (2003), \doi{https://doi.org/10.1103/PhysRevB.68.214406}.
\bibitem{Ising_CFT} J. Surace, L. Tagliacozzo, and E. Tonni, {\it Operator content of entanglement spectra in the transverse field Ising chain after global quenches}, Physical Review B {\bf 101}, 241107(R) (2020), \doi{https://doi.org/10.1103/PhysRevB.101.241107}.
\bibitem{Ising_mixedfield_CFT} O. A. Castro-Alvaredo, M. Lencs\'{e}s, I. M. Sz\'{e}cs\'{e}nyi, and J. Viti, {\it Entanglement Oscillations near a Quantum Critical Point}, Physical Review Letters {\bf 124}, 230601 (2020), \doi{https://doi.org/10.1103/PhysRevLett.124.230601}. 

\bibitem{FSS1} K. Harada, {\it Bayesian inference in the scaling analysis of critical phenomena}, Physical Review E {\bf 84}, 056704 (2011), \doi{https://doi.org/10.1103/PhysRevE.84.056704}.
\bibitem{FSS2} K. Harada, {\it Kernel method for corrections to scaling}, Physical Review E {\bf 92}, 012106 (2015), \doi{https://doi.org/10.1103/PhysRevE.92.012106}.

\bibitem{J1J2} S. Jalal and B. Kumar, {\it Edge modes in a frustrated quantum Ising chain}, Physical Review B {\bf 90}, 184416 (2014), \doi{https://doi.org/10.1103/PhysRevB.90.184416}.

\bibitem{Ryd_XXZ1} S. Geier, N. Thaicharoen, C. Hainaut, T. Franz, A. Salzinger, A. Tebben, D. Grimshandl, G. Z\"{u}rn, and M. Weidem\"{u}ller, {\it Floquet Hamiltonian engineering of an isolated many-body spin system}, Science {\bf 374}, 1149 (2021), \doi{DOI: 10.1126/science.abd9547}. 
\bibitem{Ryd_XXZ2} P. Scholl, H. J. Williams, G. Bornet, F. Wallner, D. Barredo, L. Henriet, A. Signoles, C. Hainaut, T. Franz, S. Geier, A. Tebben, A. Salzinger, G. Z\"{u}rn, T. Lahaye, M. Weidem\"{u}ller, and A. Browaeys, { \it Microwave Engineering of Programmable $XXZ$ Hamiltonians in Arrays of Rydberg Atoms}, PRX Quantum {\bf 3}, 020303 (2022), \doi{https://doi.org/10.1103/PRXQuantum.3.020303}.
\bibitem{Ryd_XXZ3} T. Franz, S. Geier, C. Hainaut, A. Braemer, N. Thaicharoen, M. Hornung, E. Braun, M. G\"{a}rttner, G. Z\"{u}rn, and M. Weidem\"{u}ller, {\it Observation of anisotropy-independent magnetization dynamics in spatially disordered Heisenberg spin systems}, Physical Review Research {\bf  6}, 033131 (2024),
\doi{https://doi.org/10.1103/PhysRevResearch.6.033131}.
\bibitem{Ryd_XXZ4} T. Franz, S. Geier, C. Hainaut, A. Signoles, N. Thaicharoen, A. Tebben, A. Salzinger, A. Braemer, M. Gärttner, G. Z\"{u}rn, and M. Weidem\"{u}ller, {\it Emergent pair localization in a many-body quantum spin system}, arXiv:2207.14216, \doi{https://doi.org/10.48550/arXiv.2207.14216}.
\bibitem{Ryd_XXZ5} G. Emperauger, M. Qiao, G. Bornet, C. Chen, R. Martin, Y. T. Chew, B. G\'{e}ly, L. Klein, D. Barredo, A. Browaeys, and T. Lahaye, {\it Benchmarking direct and indirect dipolar spinexchange interactions between two Rydberg atoms}, Physical Review A {\bf 111}, 062806 (2025), \doi{https://doi.org/10.1103/PhysRevA.111.062806}
\bibitem{Ryd_XXZ6} A. Signoles, T. Franz, R. Ferracini Alves, M. G\"{a}rttner, S. Whitlock, G. Z\"{u}rn, and M. Weidem\"{u}ller, {\it Glassy dynamics in a disordered Heisenberg quantum spin system}, Physical Review X {\bf 11}, 011011 (2021), \doi{https://doi.org/10.1103/PhysRevX.11.011011}.
\bibitem{Ryd_XXZ7} L.-M. Steinert, P. Osterholz, R. Eberhard, L. Festa, N. Lorenz, Z. Chen, A. Trautmann, and C. Gross, {\it Spatially Tunable Spin Interactions in Neutral Atom Arrays}, Physical Review Letters {\bf 130}, 243001 (2023), \doi{https://doi.org/10.1103/PhysRevLett.130.243001}.
\bibitem{Ryd_XXZ8} M. Kunimi and T. Tomita, {\it Proposal for realizing Heisenberg-type quantum-spin models in Rydberg-atom quantum simulators}, Physical Review A {\bf 112}, L051301 (2025), \doi{https://doi.org/10.1103/c97b-my2w}.

\bibitem{AOD1} D. Barredo, S. de L\'{e}s\'{e}leuc, V. Lienhard, T. Lahaye, and A. Browaeys, {\it An atom-by-atom assembler of defect-free arbitrary two-dimensional atomic arrays}, Science {\bf 354}, 1021 (2016), \doi{10.1126/science.aah3778}.
\bibitem{AOD2} M. Endres, H. Bernien, A. Keesling, H. Levine, E. R. Anschuetz, A. Krajenbrink, C. Senko, V. Vuletic, M. Greiner, and M. D. Lukin, {\it Atom-by-atom assembly of defect-free one-dimensional cold atom arrays}, Science {\bf 354}, 1024 (2016), \doi{10.1126/science.aah3752}. 
\bibitem{SLM} H. Kim, W. Lee, H. Lee, H. Jo, Y. Song, and J. Ahn, {\it In situ single-atom array synthesis using dynamic holographic optical tweezers}, Nature Communications {\bf 7}, 13317 (2016), \doi{https://doi.org/10.1038/ncomms13317}.

\bibitem{Ryd_local1} K. Kim, F. Yang, K. Mølmer, and J. Ahn, {\it Realization of an Extremely Anisotropic Heisenberg Magnet in Rydberg Atom Arrays}, Physical Review X {\bf14}, 011025 (2024), \doi{https://doi.org/10.1103/PhysRevX.14.011025}.
\bibitem{Ryd_local2} G. Bornet, G. Emperauger, C. Chen, F. Machado, S. Chern, L. Leclerc, B. G\'{e}ly, Y. T. Chew, D. Barredo, T. Lahaye, N. Y. Yao, and A. Browaeys, {\it Enhancing a Many-Body Dipolar Rydberg Tweezer Array with Arbitrary Local Controls}, Physical Review Letters {\bf 132}, 263601 (2024), \doi{https://doi.org/10.1103/PhysRevLett.132.263601}.


\bibitem{2D_future1} Y. D. Liao, G. Pan, W. Jiang, Y. Qi, and Z. Y. Meng, {\it 
The teaching from entanglement: 2D SU(2) antiferromagnet to valence bond solid deconfined quantum critical points are not conformal}, arXiv:2302.11742, \doi{https://doi.org/10.48550/arXiv.2302.11742}.
\end{thebibliography}
\end{document}